\documentclass[fleqn,10pt]{wlscirep}
\usepackage{graphicx}
\usepackage{dcolumn}
\usepackage{bm}
\usepackage{amsmath}
\usepackage{amsfonts}
\usepackage{amssymb}
\usepackage{color}
\usepackage{soul}
\usepackage{epstopdf}
\usepackage{epsfig}

\def\be{\begin{equation}}
\def\ee{\end{equation}}
\def\ba{\begin{eqnarray}}
\def\ea{\end{eqnarray}}
\def\nn{\nonumber}
\newcommand{\mb}[1]{\mathbf{#1}}

\newcommand{\de}[1]{\textcolor{black}{#1}}

\newcommand{\kk}[1]{\textcolor{black}{#1}}

\newcommand{\jb}[1]{\textcolor{black}{#1}}
\newcommand{\jbt}[1]{\textcolor{black}{#1}}

\begin{document}

\title{Coupled multiple-mode theory for  $s_\pm$ pairing mechanism in iron based superconductors}

\author[1,*]{M. N. Kiselev}
\author[2,+]{D.V. Efremov}
\author[2]{S. L. Drechsler}
\author[2]{Jeroen van den Brink}
\author[3]{K. Kikoin}

\affil[1]{The Abdus Salam International Centre for Theoretical
Physics, Strada Costiera 11, I-34151 Trieste, Italy}
\affil[2]{Institute for Theoretical Solid State Physics, IFW Dresden, Germany}
\affil[3]{School of Physics and Astronomy, Tel Aviv University, 69978 Tel Aviv, Israel}

\affil[*]{mkiselev@ictp.it}
\affil[+]{d.efremov@ifw-dresden.de}


\begin{abstract}
We investigate the interplay between the magnetic and the superconducting degrees of freedom in unconventional multi-band superconductors such as iron pnictides. For this purpose a dynamical mode-mode coupling theory is developed based on the coupled Bethe-Salpeter equations. In order to investigate the region of the phase diagram not too far from the tetracritical point where the magnetic spin density wave, (SDW) and superconducting (SC) transition temperatures coincide, we also construct a Ginzburg-Landau functional including both SC and SDW fluctuations in a critical region above the transition temperatures.
\kk{The fluctuation corrections \jbt{tend to suppress the} magnetic transition, but \jbt{in the superconducting channel the intraband and interband contribution of the} fluctuations nearly compensate each other.}
\end{abstract}

\flushbottom
\maketitle

\section*{Introduction}

The rapidly extending realm of high-$T_c$ superconductors has been enriched recently by a new class of materials,
so called iron based superconductors (FeSC). \cite{Johnston10,PaGree10,AnBoe10,HiKoMa11,Chubuk11} Like many other  high-$T_c$ materials,
these compounds crystallize in strongly anisotropic \jb{lattices: one can identify quasi-two-dimensional subsystems which contain the electrons that are subject} to Cooper pairing. In \jb{FeSCs these planes consist} of square pyramids with alternatively oriented apex vertices that are occupied by pnictogen ions and a square base formed by iron ions.
A specific feature of the electronic structure of FeSC is the multipocket Fermi surface, mostly semi-metallic
(with both hole and electron pockets), although some superconducting materials possess either only electron \cite{Mou11,Liu} or only hole pockets \cite{holep}.  Another
characteristic feature of FeSC is the interplay between different electronic instabilities of the pristine normal metal phases.
Most of these materials are unstable against spin-density wave (SDW) type itinerant antiferromagnetism, and the phase
diagrams of doped compounds contains domains of superconductor (SC) and SDW ordering, the latter sometimes being
accompanied by structural phase transitions with \jb{a potential for} orbital ordering. \cite{Orbit1,avchi,Orbit2,Orbit3} This means that any consistent theory of superconductivity should take into account the
strong interplay between electronic and magnetic instabilities in presence of  trends to orbital ordering and soft lattice displacement modes {\color{black} (see another more complex approach for description of d-electrons in iron pnictides in [\citeonline{TetGor2013}])}.
\textcolor{black}{Such a more general  approach to consider not simply a single instability of a
 "normal state"  phase with respect to superconductivity
but  instead to treat on equal footing the competition (including also their coexistence) of various orderings is a generic problem for  many
non-standard superconductors and related phases \cite{Carlson,Kivelson}. Even the standard  case of strong electron-phonon interaction mediated SC and a Fermi surface derived instability
requires strictly speaking the account of possible lattice instabilities and of the related anharmonicities
reducing the strength of the  former. For the sake of simplicity here we will only focus on the interplay of a specific SDW and SC
\jb{-} a generalization to the case of  higher multi-component instabilities is straightforward.
}

\jb{A specific motivation to consider and develop a multi-mode theory is provided by the experimental observation that in FeSCs electron or hole doping of pristine magnetic materials is crucial for suppression of itinerant SDW magnetism and formation of superconducting state. Apart from that,} doping by isovalent impurities or a noticeable concentration of intrinsic defects (vacancies) can also strongly modify their phase diagram. It was noticed in this context that conventional understanding of the role of non-magnetic and magnetic impurities in formation of Cooper pairs in BCS superconductors formulated in classical papers by A.A. Abrikosov, L.P. Gor'kov and P.W. Anderson (see, e.g., \cite{AG59} and \cite{Sadovskii2006}) should be revisited and modified for these multiband superconductors because interband coupling plays crucial role in formation of superconducting order in these nearly nested, quasi 2D materials. Both nonmagnetic and magnetic scattering essentially influence $T_{\rm c}$ - in particular, interband scattering due to non-magnetic impurities is destructive for $s_\pm$ superconductivity, which is considered to be realized in most of FeSC \cite{GolMa97,SenKon08,Efrem11,EGDo13,Kontani13}
and magnetic interband scattering  on the contrary stabilizes the $s_\pm$ coupling mechanism.\cite{GolMa97,Efrem14} The kinematics of this stabilization mechanism is similar to that in so called $\pi$-junctions in heterostructures superconductor/insulator/superconductor, where the Josephson tunnelling through a barrier with paramagnetic impurities is accompanied with the sign change of the superconducting order parameter $\Delta$. \cite{Bulaev77}
Another remarkable response to doping occurs when nominally nonmagnetic defects such as As vacancies \cite{Fuchs08,Fuchs09,Grine11} or Ru substitutions\cite{Sanna13} \jb{are introduced} in 1111 materials. These impurities trigger a strong paramagnetic response {\color{black}(see \cite{KKSMB} for explanation of this effect),} and their observed modification of $T_{\rm c}$ does not fit with available theories.

To this end we introduce and develop a multi-mode theory that takes into account the fact that the superconductivity in these materials arises as a result of competition of at least two coupling mechanisms: superconducting pairing with the scalar order parameter $\Delta$  and the spin density wave \de{ordering} with vector order parameter $\bf m$ (the magnetic moment). These modes are coupled by the interband electron-electron interaction, and we consider the effect of impurity scattering on this coupling mechanism.

The paper is written as follows. We start with the derivation of
a system of Bethe-Salpeter (BS) equations containing coupled SDW and Cooper channels for temperatures above the magnetic and superconducting transition temperatures in the next section. 
 Based on BS equations we show that dynamical fluctuations play an important role close to the tetracritical point and must be taken into account for the correct description of the tetracritical region.

Furthermore we derive the Landau-Ginzburg (LG) functional and calculate fluctuation corrections to the phase transition temperatures to superconductor and SDW phases ($T_{\rm c}$ and $T_{\rm s}$, respectively).
Detailed derivation of BS and LG equations is given in the Section "Methods". 

Where needed, we will be specific and concentrate on the multi-valley semi-metallic FeSCs from the 1111 (\textit{Re}FeAsO) and 122
(\textit{Ae}Fe$_2$As$_2$)
families under electron and hole doping (\textit{Re} and \textit{Ae} stay for the rare-earth and alkaline-earth elements) respectively.

\section*{Multimode approach to SDW and BCS instabilities}
\label{sec:derivation}

In this Section we formulate a mode coupling approach for multiband ferropnictide superconductors with nearly nested hole and electron pockets of Fermi surface.
 Although a complete consensus is lacking on the type of superconductor pairing in these system, it appears that the experimental arguments in favor of $s_\pm$ mechanism are rather solid. This type of ordering was proposed originally
for superconductor-excitonic instability in semimetals \cite{ArSon73} and has been reformulated for superconductor pairing
in FeSCs in Refs. \cite{MSJD08,Kuroku08} (for a review see  \cite{Chubuk11}).
It is important to note that mode-mode coupling is an inherent constituent of $s_\pm$ pairing \cite{CheEE08,MaChub10,Chubukov2016}:
in the two-dimensional systems with nesting conditions between electron and hole pockets
an interplay between the SDW and Cooper modes results in additional coupling within each channel induced by all other channels. This theory is based on a
renormalization group (RG) approach, which implies logarithmic renormalization of the relevant vertices. It was shown that the Cooper and SDW decouple at the scale of Fermi energy and further flow goes independently in both channels.
The natural limitation of such approach is the demand of perfect nesting in SDW channel. In real FeSC systems the
nesting conditions are satisfied only approximately. It leads to the suppression of the SDW channel at some energy.

The general approach to deal with mode-mode coupling that we introduce here is based on the mode coupling theory of critical phenomena. We start from the
"high temperature" region, where the critical fluctuations are already well developed, while long-range SC and SDW
order are not yet established. The corresponding vertex parts
$\Gamma_{\rm sc}(q,\Omega,T)$  and $\Gamma_{\rm sdw}({\bf Q+q},\Omega,T)$, are represented at this temperature by the {\color{black} fluctuating} modes $ D_{\rm sc} (q,\Omega)$
and $D_{\rm sdw} (q,\Omega)$ respectively \cite{Varla, Var2011}:
\begin{eqnarray}\label{flusc}
&& \nu \Gamma_{\rm sc}(q,\Omega,T) \to D_{\rm sc} (q,\Omega) =
\frac{\nu \Gamma_{\rm sc}^{(0)}}{-\frac{i\Omega}{\gamma_{\rm sc}}+\tau_{\rm c} +c_\Delta q^2},\\
&&  \nu \Gamma_{\rm sdw}(\mathbf{Q+q},\Omega,T)\to D_{\rm sdw} (q,\Omega)  =
\frac{\nu \Gamma_{\rm sdw}^{(0)}}{-\frac{i\Omega}{\gamma_{\rm sdw}}+\tau_{\rm s} +c_m q^2}, \nonumber
\end{eqnarray}
with critical parameters
\begin{eqnarray}  \tau_{\rm c}=\frac{T-T_{\rm c}}{T_{\rm c}},~~
\tau_{\rm s}=\frac{T-T_{\rm s}}{T_{\rm s}},
\end{eqnarray}
and $\gamma_{\rm sc/sdw}=8T/\pi$ . Here we consider the limit of small momenta $c_{\Delta/m}q^2 \ll 1$, {\color{black} $\nu$ is a density of states at the Fermi level,  $c_\Delta=\xi_s^2$ and $c_m=\xi^2_m$
are  superconducting and magnetic coherence lengths  respectively, see Section "Ginzburg-Landau approach". 
}

As the pole $i\Omega_{\rm sc/sdw}= \gamma_{\rm sc/sdw}(\tau_{\rm c/s} +c_{\Delta/m}q^2)$ of the vertex part $\Gamma_{\rm sc}$ or $\Gamma_{\rm sdw}$ at $q\to 0$ tends to zero, the corresponding instability results in the phase transition.
When approaching the transition temperature $T_{\rm c}$ or $T_{\rm s}$, the corresponding vertex part diverges in the static limit:
\begin{equation}\label{diverg}
\Gamma_{\rm sc}(0,0,T)=\Gamma_{\rm sc}^{(0)}\tau_{\rm c}^{-1}, ~\Gamma_{\rm sdw}(Q,0,T)=\Gamma_{\rm sdw}^{(0)}\tau_{\rm s}^{-1}.
\end{equation}

\begin{figure}[t]
\begin{center}
  \includegraphics[width=.3\columnwidth,angle=0]{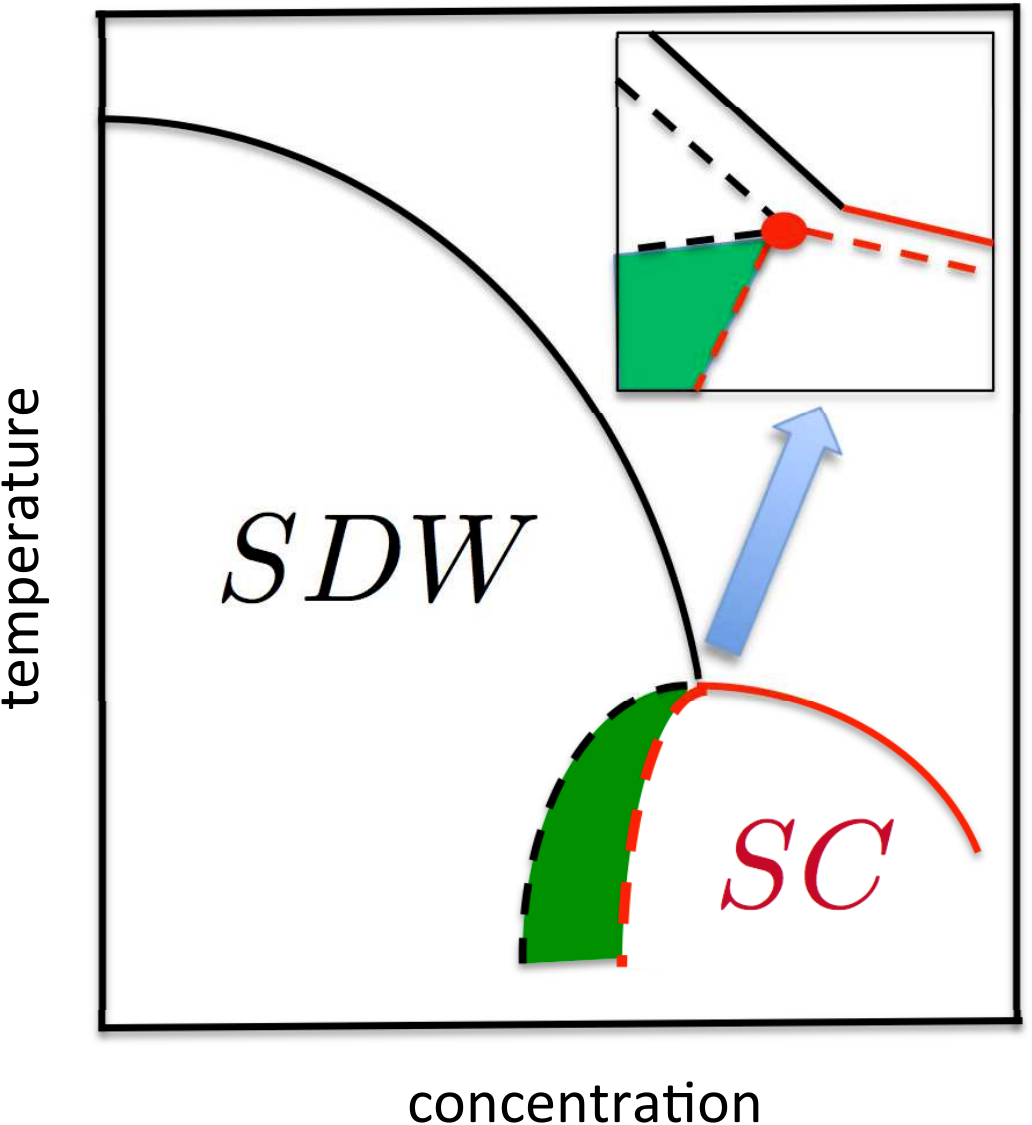}
\end{center}
\caption{(Color online) Phase diagram of competing SDW and SC states as a function of the carrier concentration. The shaded area denotes the coexistence region of $s_\pm$ superconductivity and the SDW.
The critical temperatures of SDW and SC transitions coincide in the tetracritical point. The insert shows zoomed in part of the phase diagram around the mean-field tetracritical point: the new tetracritical point
(filled circle) is shifted due to fluctuations computed in the multi-mode coupling theory (dashed lines).}
\label{fig_2}
\end{figure}

 The mode coupling approach is valid in the vicinity of the tetracritical point on the phase diagram shown in Fig. \ref{fig_2} where the two temperatures $T_{\rm c}$ and $T_{\rm s}$ are comparable. Then the system of equations for the vertex parts at small $\Omega$ and $q$ can be derived and solved. These are the Bethe-Salpeter equations for the most relevant vertices. As will be shown below, this system contains at least four bare vertices which should be taken into account. The divergences (\ref{diverg}) provide us with a criterion of selecting the corresponding diagrams. Namely, the terms containing polarization operators in the electron-hole channel with transmitted momentum close to Q, and the electron-electron polarization operators of Cooper type should be collected in all orders. As a result, the system of Bethe - Salpeter equations acquires a parquet-like structure, and the solution of the mode-coupling equations obtained at $T > T_{\rm c}$ may be mapped on the mean-field solutions of renomalization group (RG) equations below $T_{\rm c}$ \cite{CheEE08,MaChub10}.

We assume that the superconducting transition temperature $T_c$ is imposed on the system by a dominant $s_\pm$
pairing. In this Ansatz the interband electron-hole Coulomb interaction is the main source of Cooper pairing, so that a weak BCS attractive interaction in the electron and hole pockets would result in SC instabilities with
$T_{\rm c}^{(e,h)} \ll T_{\rm c}$.
\medskip

\subsection*{Bethe-Salpeter equations}
\begin{figure}[t]
\begin{center}
  \includegraphics[width=.8\columnwidth,angle=0]{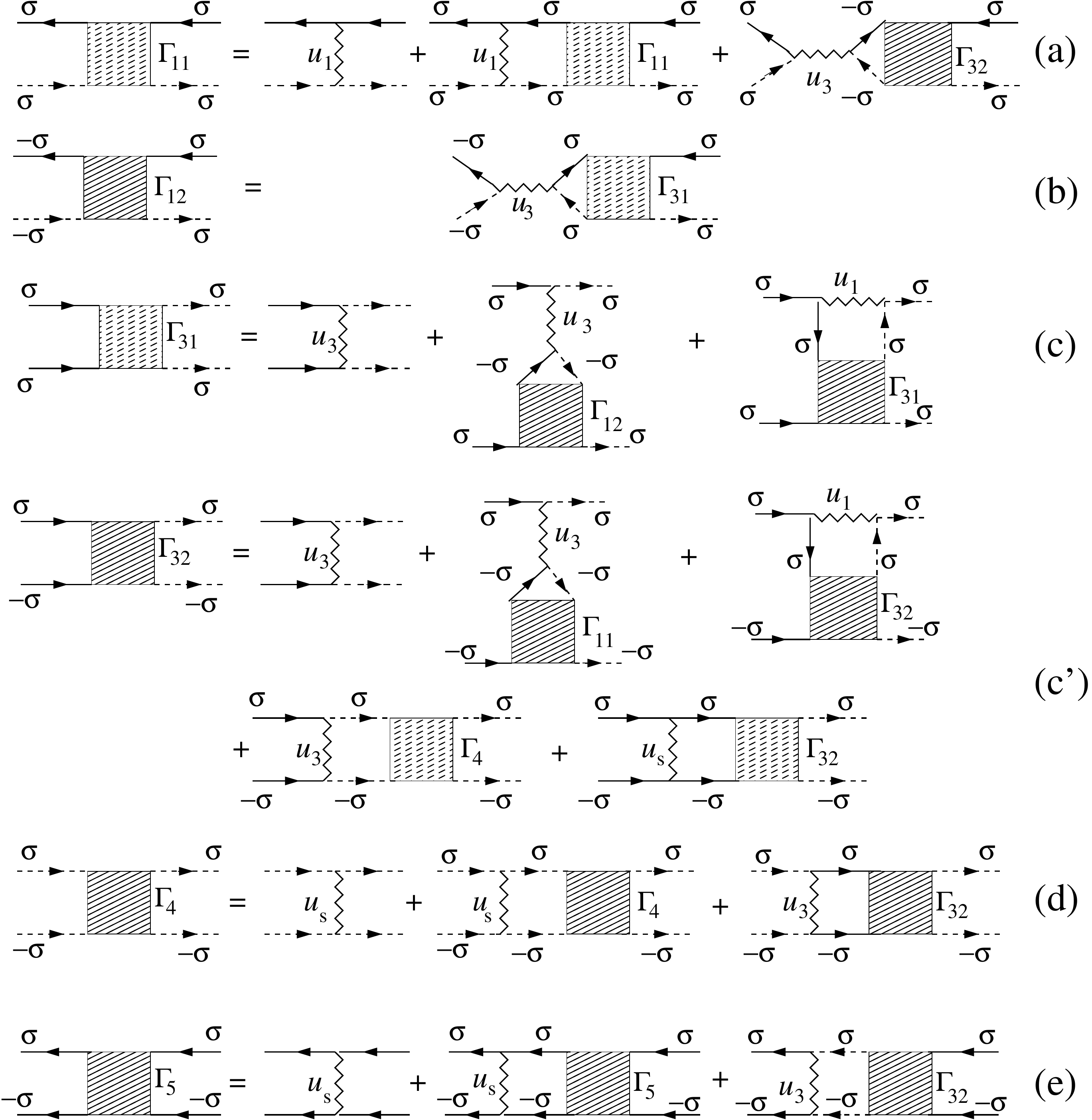}
\end{center}
\caption{
Diagrams for the system of Bethe-Salpeter equations. Solid and dashed lines stand for hole and electron propagators, respectively. The SDW polarization loop $\Pi_1^{\sigma\sigma}$ contains one electron and one hole bare Green function. The Cooper loops $\Pi_s$ contain two electron or two hole bare propagators, respectively.
}
\label{fig_1}
\end{figure}

{\color{black} Following the Ansatz formulated above (see details in the Section "Methods" \ref{sec:apa}), we derive  the system of BS equations for the relevant vertex parts
$\Gamma_i$ (see Fig. \ref{fig_1}). The starting Hamiltonian is:
\begin{eqnarray}\label{fullHamiltonian}
H &=& \sum_{\mathbf{k}, \sigma} \left[(\epsilon^e_{\mathbf{k}}-\mu) f^\dagger_{\sigma \mathbf{k}}f_{\sigma \mathbf{k}} + (\epsilon^h_{\mathbf{k}}-\mu)c^\dagger_{\sigma \mathbf{k}}c_{\sigma \mathbf{k}}\right]
  + \sum_{\mathbf{k_1+k_2= k_3+k_4 , \sigma  \sigma'}} \left[u_1 c^\dagger_{\sigma \mathbf{k_1}} f^\dagger_{\sigma' \mathbf{k_2}}f_{\sigma' \mathbf{k_3}} c_{\sigma \mathbf{k_4}}
  + u_2  c^\dagger_{\sigma \mathbf{k_1}} f^\dagger_{\sigma' \mathbf{k_2}}c_{\sigma' \mathbf{k_3}} f_{\sigma \mathbf{k_4}}\right] \nonumber \\
  &+&   \sum_{\mathbf{k_1+k_2= k_3+k_4, \sigma  \sigma'}}\left[\frac{u_3}{2} c^\dagger_{\sigma \mathbf{k_1}} c^\dagger_{\sigma' \mathbf{k_2}}f_{\sigma' \mathbf{k_3}} f_{\sigma \mathbf{k_4}}
  + \frac{u_4}{2}  c^\dagger_{\sigma \mathbf{k_1}} c^\dagger_{\sigma' \mathbf{k_2}}c_{\sigma' \mathbf{k_3}} c_{\sigma \mathbf{k_4}}
  + \frac{u_5}{2}  f^\dagger_{\sigma \mathbf{k_1}} f^\dagger_{\sigma' \mathbf{k_2}}f_{\sigma' \mathbf{k_3}} f_{\sigma \mathbf{k_4}} + h.c. \right],
\end{eqnarray}
where $f_{\sigma \mathbf{k} }$,  $c_{\sigma \mathbf{k} }$ stand for annihilation operator in electron and hole bands;  $\epsilon^{e/h}_{\mathbf{k}}$ and $\mu$ are dispersions and chemical potential, respectively.

Based on the results of RG calculations \cite{CheEE08,MaChub10}, we choose four vertices relevant to the $s_\pm$ coupling (the vertices $u_1, u_3, u_4=u_s, u_5=u_s$). The vertex $u_2$ which has been shown to be irrelevant\cite{MaChub10} is excluded.}

The vertices $\Gamma_{1i}$ with $i=1,2$ (see Fig. \ref{fig_1}) describe the interactions in the density-wave block,
the vertices $\Gamma_{4,5}$ describe the singlet Cooper pairing in the electron and hole pockets, respectively,
and the vertex part $\Gamma_3$ includes the interactions responsible for the interband (\textit{e-h})
Cooper pairing. The system of Bethe--Salpeter equations may be written in the symmetric form using a matrix $\hat \Lambda$:
\begin{equation}\label{besa}
(1- \hat \Lambda)\hat  \Gamma = \hat u.
\end{equation}
The matrix
$(1-\hat \Lambda)$ is the secular matrix and
\begin{equation}\label{ma1}
{\hat u}=
\left(
\begin{array}{c}
 u^{}_{1}\\
 u^{\sigma\bar\sigma}_{3} \\
 u^{\sigma\sigma}_{3} \\
 u_s \\
 u_s
\end{array}
\right), ~~~
{\hat \Gamma}=
\left(
\begin{array}{c}
 \Gamma^{}_{11}\\
 \Gamma^{}_{31} \\
 \Gamma^{}_{32} \\
 \Gamma_4\\
 \Gamma_5
\end{array}
\right),
\end{equation}
The secular equation for this system of Bethe-Salpeter equations for the modes
\begin{eqnarray}\label{propa}
D_1^{-1}&=& 1 - u^{}_1\Pi_1^{\sigma\sigma}, \nonumber \\
D_{31}^{-1}&=&  1 - u_1 \Pi_1^{\sigma\sigma}  - u_3^2 \Pi_1^{\sigma \sigma} \Pi_1^{\bar \sigma \bar \sigma},\nonumber\\
D_{32}^{-1}&=&  1 -  u^{}_1\Pi_1^{\sigma\sigma} - u_s\left(\Pi_{se}+\Pi_{sh}\right)/2,\nonumber \\
D_4^{-1}&=& 1-u_s\Pi_{se}, \nonumber \\
D_5^{-1}&=& 1 -u_s\Pi_{sh},
\label{eq.modes}
\end{eqnarray}
is $ \det  (1-\hat \Lambda) =0$.
In these notations both coupling constants $u_i=\nu U_i$ and the polarization loops
$\Pi_i(q=0,\omega=0) = \ln (W/T)$ are dimensionless (here $W$ is the bandwidth).
{\color{black}
We neglect below the difference between the band dispersion in the electron and hole pockets, when calculating the polarization operators $\Pi_s=\Pi_{se} = \Pi_{sh}$. \jbt{One should note that his approximation does not imply that we rely on perfect nesting. In electron doped materials the strict nesting conditions are of course} violated, but the small difference between the electron and hole polarization operators is insufficient for our theory because two contributions are summed in the critical mode $D_{32}$  which \jbt{is central to} our scenario. The criticality in this mode is dominated by the term $u_1\Pi_1$.}
We use the sign convention that both $\Pi_1$ and $\Pi_s$ are positive. Under this convention $u_s>0$ is attractive in the Cooper channel and  $u_1>0$
facilitates the instability in the SDW channel.

In the explicit form the secular equation reads:
\begin{equation}\label{secul}
{\rm det}(1- \hat \Lambda) = \left|
 \begin{array}{ccccc}
 D_1^{-1}(p,\omega)  &  0 & -u_3\Pi^{\bar\sigma\bar\sigma}_1 & 0& 0\\
 0 & D_{31}^{-1}(p,\omega)  & 0 & 0 & 0 \\
-u_3\Pi_1^{\bar\sigma\bar\sigma} & 0  & D^{-1}_{32} (p,\omega)  &-u_3\Pi_s&    -u_3\Pi_s\\
0 & 0 &-u_3\Pi_s  &  D^{-1}_4(p,\omega)  &0 \\
 0 & 0 & -u_3\Pi_s & 0 & D^{-1}_5(p,\omega)\\
 \end{array}
\right|=0.
\end{equation}

 One can see that in this approximation the ${\cal T}_{\pm 1}$ triplet $\Gamma_{31}$-channel described by the second row of the secular equation  decouples from the rest, which corresponds to the ${\cal T}_0$ - SDW and superconducting channels. The $D_{32}$-mode plays a very particular role. On the one hand it contributes to the density wave channels (both CDW and SDW). On the other hand it strongly affects the superconducting channel.   In the approximation used in the present paper the SDW and CDW are degenerate. The interband interaction $u_2>0$ with momentum transfer $\mathbf{Q}$ lifts out the degeneracy favouring the SDW transition.

We will consider the part of the  phase diagram concentration-temperature ($c-T$) close to the  point of the degeneracy of the $s_\pm$ and SDW channels (i.e. the tetracritical point) shown in Fig. \ref{fig_2}. In this region the SC instability takes place in the presence of critical SDW fluctuations.
{\color{black} Above $T_{\rm s}$ the fluctuation modes  arise at momenta $\mathbf{p} =\mathbf{Q}+\mathbf{q}$ close to the nesting vector ${\bf Q}$ connecting the $\Gamma$ and $X$ points in the Brillouin zone and at small $\omega \to 0$.}

We assume that the two Cooper propagators $D_4$ and $D_5$ are far from any divergence, namely  $u_s\Pi_s\ll 1$ at these temperatures (in case of a dominant interband pairing mechanism as in the pronounced $s_\pm$ case adopted here, the purely intraband Cooper instabilities
develop at temperatures much less than the actual $T_{\rm c}$). In this approximation the vertices are real, and the channels 4,5 are represented by a single row and column.

\subsection*{Away from the tetracritical point}

We consider first the case when the temperature {\color{black}is less than} the Fermi-energy $T <  \epsilon_F$ and the doping $c$ is away
from the tetracritical point. Then the divergence of the vertices is strongly peaked at particular momenta.
For instance, putting  $u_3\to 0$ in the  integral equation for the
$\Gamma_{11}(\mathbf{p}_1,\mathbf{p}_2,\mathbf{p}_3,\mathbf{p}_4)$ (see Fig. \ref{fig_1} a),
one immediately gets that this vertex is divergent only for $\mathbf{p}_1 + \mathbf{p}_2 = \mathbf{Q}$.
A similar analysis shows that  $\Gamma_{32}$ diverges  for $\mathbf{p}_1+\mathbf{p}_2 \to 0$ and
$\mathbf{p}_1-\mathbf{p}_3\to \mathbf{Q}$.
The splitting of momenta, at which the vertices diverge, decouples the matrix Eq.(\ref{besa})
into the density-wave  and Cooper channels.
To study the properties of the vertex functions in the vicinity of their singularities we introduce
$\Gamma_{11}(\mathbf{p}_1,\mathbf{p}_2,\mathbf{p}_3,\mathbf{p}_4) \approx
\Gamma^{(a)}_{11}(\mathbf{p}_1 + \mathbf{p}_2 - \mathbf{Q})+
\Gamma^{(b)}_{11}( \mathbf{p}_1 - \mathbf{p}_3 + \mathbf{Q})$, where $\Gamma^{(a)}_{11}(\mathbf{\tilde{k}})$
and $\Gamma^{(b)}_{11}(\mathbf{p})$ have  poles at $\mathbf{k},\mathbf{p} \to 0$ correspondingly. For the vertices
$\Gamma_{32}$ and $\Gamma_4$ {\color{black} we use} the same decomposition: $\Gamma_{32}(\mathbf{p}_1,\mathbf{p}_2,\mathbf{p}_3,\mathbf{p}_4) \approx
\Gamma^{(a)}_{32}(\mathbf{p}_1 + \mathbf{p}_2 )+
\Gamma^{(b)}_{32}( \mathbf{p}_1 - \mathbf{p}_3 + \mathbf{Q})$
 and $\Gamma_{4}(\mathbf{p}_1,\mathbf{p}_2,\mathbf{p}_3,\mathbf{p}_4) \approx
\Gamma^{(a)}_{4}(\mathbf{p}_1 + \mathbf{p}_2 )+
\Gamma^{(b)}_{4}( \mathbf{p}_1 - \mathbf{p}_3 + \mathbf{Q})$.

To identify the corresponding transition temperatures we consider the Bethe-Salpeter equations in the static limit.
For small momenta $\mb{p},\tilde{\mb{k}},\mb{q}$ the system  Eq.(\ref{besa})  splits up in the logarithmic approximation (see Section "Methods" \ref{sec:apa}) into cooper channel:
\begin{equation}
\Gamma^{(a)}_{4}(\mb{k}) \pm \Gamma_{32}^{(a)}(\mb{k}) = \frac{u_s\pm u_3}{1- (u_s\pm u_3)\Pi_s(\mb{k})},
\label{eq.gamma4+gamma32}
\end{equation}
and density-wave channel
\be
\Gamma^{(a)}_{11}(\mb{k}) \pm \Gamma_{32}^{(b)}(\mb{k}) = \frac{u_1\pm u_3}{1- (u_1\pm u_3)\Pi_1(\mb{k})}.
\label{eq.gamma11+gamma32}
\ee
With these vertices we now calculate the susceptibility in the cooper and density-wave channels.
In the particle-particle $s_\pm$ and $s_{++}$ channel drawn in Fig.~\ref{fig_3} we obtain
\be
\chi^{SC}_{\pm, ++}(\mb{k},T) =  \Pi_s(\mb{k},T)+\Pi_s^2(\mb{k},T)(\Gamma_{4}(\mb{k},T) \pm \Gamma_{32}(\mb{k},T)).
\label{eq.chi_SC1}
\ee
Substituting Eq. (\ref{eq.gamma4+gamma32}) into Eq. (\ref{eq.chi_SC1}), we get:
\be
\chi^{SC}_{\pm, ++}(\mb{k},T) = \frac{\Pi_s(\mb{k},T)}{1 - (u_s\pm u_3)\Pi_s(\mb{k},T)}.
\label{eq.chi_SC2}
\ee
This equation has the typical pole structure for the superconducting susceptibility in the vicinity to the transition temperature.
 The susceptibility $\chi^{SC}_{\pm, ++}(\mb{k}=0,T)$ increases with decreasing temperature and diverges at the
critical transition temperature.
This happens  under condition $1 - (u_s\pm u_3)\Pi_s(\mb{k}=0,T=T_c) =0$.  The two solutions correspond to $s_{++}$ and $s_{\pm}$ superconducting
order parameter.
For CDW and SDW channels we get:

\begin{figure}[t]
\begin{center}
  \includegraphics[width=.3\columnwidth,angle=0]{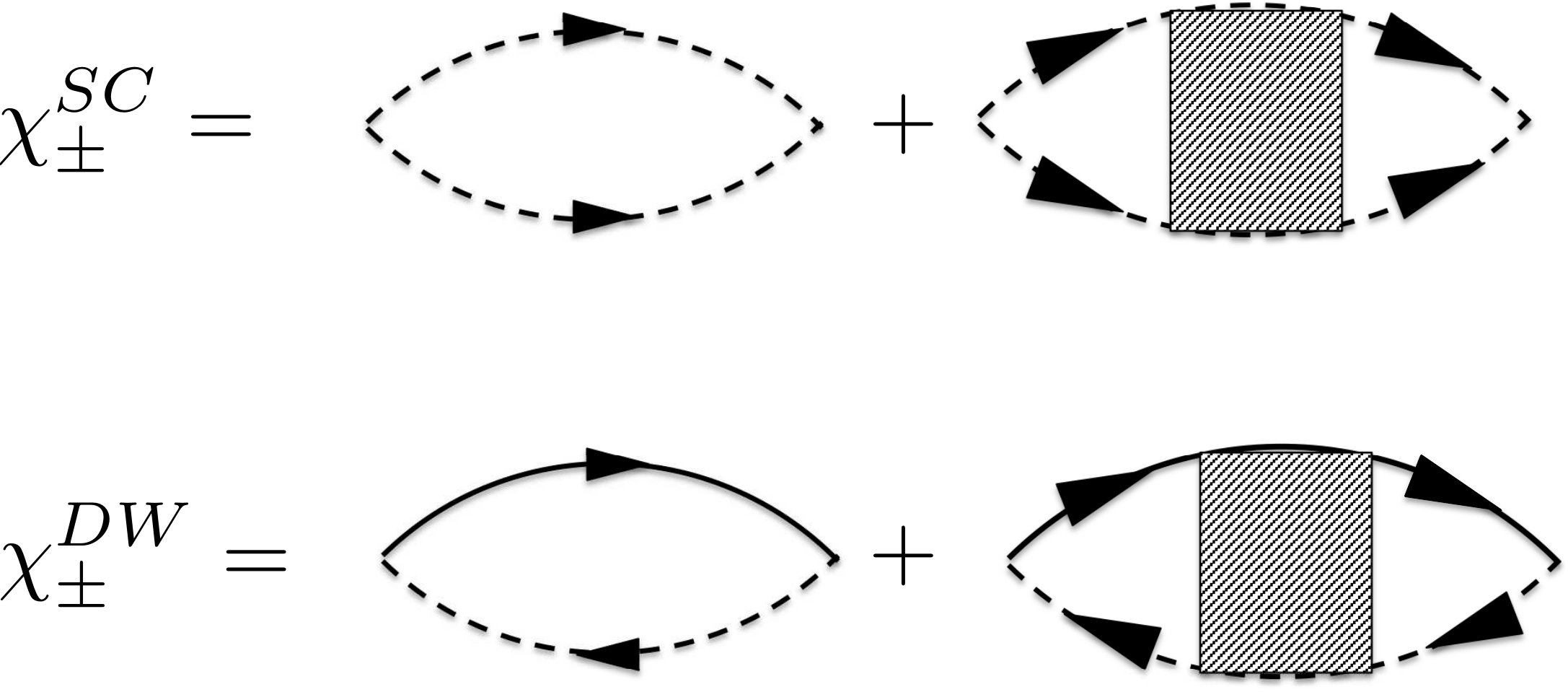}
\end{center}
\caption{Cooper and density waves susceptibilities. {\color{black} The shaded boxes denote} the corresponding combination of the vertices (see the text for details).}
\label{fig_3}
\end{figure}

\ba
\chi^{DW}_{\pm}(k,T) &=& 
\frac{\Pi_1(\mb{k},T)}{1-(u_1\pm u_3)\Pi_1(\mb{k},T)}.
\label{eq.chi_SD}
\ea
The SDW transition is determined by $1-(u_1+u_3)\Pi_{1}(q=0,T_c) =0$.
One can define a tetracritical point, when both susceptibilities diverge.
It happens under the condition $(u_s+u_3)\Pi_s(\mb{k}=0,T_c)=(u_1+u_3)\Pi_1(\mb{k},T_c)\to 1 $. In this case and in the vicinity of the
tetracritical point one should find the divergence of the susceptibility taking into account the full matrix Eq.(\ref{besa}).

\subsection*{Close to the tetracritical point: dynamical multi-mode coupling theory}

Now we consider the interplay between magnetic
 and superconducting degrees of freedom  in the vicinity of the tetracritical point. In this case the approach based on the effective hydrodynamic action is much easier than the direct  solution of the BS equations. {\color{black} Besides, this approach which also accounts for the multi-point correlation functions goes beyond the conventional BS paradigm limited by two- and four- point Green's function \cite{AG59, Sadovskii2006}.} The  derivation of the effective hydrodynamic action in terms of  magnetic
$\vec m(\vec x,t)$ and superconducting $\Delta_i(\vec x,t)$ (i=e,h) dynamical fluctuations is done by integrating out the BS equations with respect to the "fast" (with the energy of the order of the bandwidth) degrees of freedom and is presented  in Section "Methods" \ref{sec:apb}.

Within our analysis we start discussing the limiting case $u_s\ll u_3$ and $u_3\Pi_s \sim 1$ corresponding to the interplay
between two fluctuating modes: one is the $s_\pm$ superconducting and another one is the SDW magnetic.
This regime is believed to be present in most of FeSC [e.g., in broadly studies doped "A-122" systems (where A = Ba ,Sr, Ca)]
The Lagrangian of the two-mode coupling theory takes the form:
\begin{eqnarray}
{\cal L}^{(2)}&=& \int d\vec x \left\{\bar\Delta_- L^{-1}_{\pm} \Delta_- +
\vec m D^{-1}_m \vec m
+ {\color{black}\frac{1}{2}} A|\Delta_-|^4+ B\vec m^4 + (C_1-C_2) |\Delta_-|^2\vec m^2\right\}
\label{l1}
\end{eqnarray}
Here the Fourier transforms of superconducting and SDW fluctuators computed on Matsubara frequencies $i\Omega_m$ {\color{black} read:
\begin{eqnarray}
&&L^{-1}_\pm(q,i\Omega_m,T)=\frac{1}{u_s+ u_3} -\Pi_s(q,i\Omega_m,T_c)=
\nu\left[\tau_c+\psi\left(\frac{1}{2}+\frac{|\Omega_m|}{4\pi T}\right)-\psi\left(\frac{1}{2}\right)-\frac{\langle(\vec v_p\cdot \vec q)\rangle_{FS}}{2(4\pi T)^2}\psi''\left(\frac{1}{2}+\frac{|\Omega_m|}{4\pi T}\right)\right],\\
&&D^{-1}_m({\bf Q+q},i\Omega_m,T)= \frac{1}{u_1+ u_3} -\Pi_1(q,i\Omega_m,T_c)=
\nu\left[\tau_s+\psi\left(\frac{1}{2}+\frac{|\Omega_m|}{4\pi T}\right)-\psi\left(\frac{1}{2}\right)-\frac{\langle(\vec v_p\cdot \vec q)\rangle_{FS}}{2(4\pi T)^2}\psi''\left(\frac{1}{2}+\frac{|\Omega_m|}{4\pi T}\right)\right]. \nonumber
\label{ld0}
\end{eqnarray}
Here $\psi(x)$ and $\psi''(x)$ are the  di-gamma function and its second derivative respectively, $\langle...\rangle_{FS}$ denotes the averaging over the Fermi surface (here a parabolic dispersion and equal mass were assumed for electron and hole pockets).
The analytic continuation of the superconducting and SDW fluctuators to the upper half-plane $i\Omega_m\to \Omega +i0^{+}$ for $\Omega \ll (T_c,T_s)$  is given by}
\begin{eqnarray}
&&L^{-1}_\pm(q,\Omega,T)=\nu\left[-\frac{i\Omega}{\gamma_{sc}} +\tau_{c} + c_\Delta q^2 \right],\\
&&D^{-1}_m({\bf Q+q},\Omega,T)= \nu\left[-\frac{i\Omega}{\gamma_{sdw}} +\tau_{s} + c_m q^2\right]. \nonumber
\label{ld1}
\end{eqnarray}
Notice that we normalize the fluctuators by the density of states to make them dimensionless.  The equations defining the coefficients $A, B, C_1$, $C_2$  and $c_{\Delta/m}$ are derived in Section "Ginzburg-Landau approach".
The static $q$ - independent part of the Lagrangian corresponds to the Landau expansion of the free energy. Inclusion of the gradient terms generalizes the Landau theory to the Ginzburg-Landau functional (see Section "Ginzburg-Landau approach"). The effective Lagrangian describes effective {\it non-linear} theory of interacting mode and therefore goes beyond {\it linear} Bethe - Salpeter approach. Hidden non-linearity of Bethe-Salpeter equations is associated with curved ("non-flat") phase space: $\Gamma_{32}$ enters both SDW and superconducting equations being singular both at small total moment/energy and small deviating from $Q$ momentum transfer.
The Lagrangian (\ref{l1}) is $U(1)\times SU(2)$ symmetric.
\begin{figure}[t]
\begin{center}
\includegraphics[width=.7\columnwidth,angle=0]{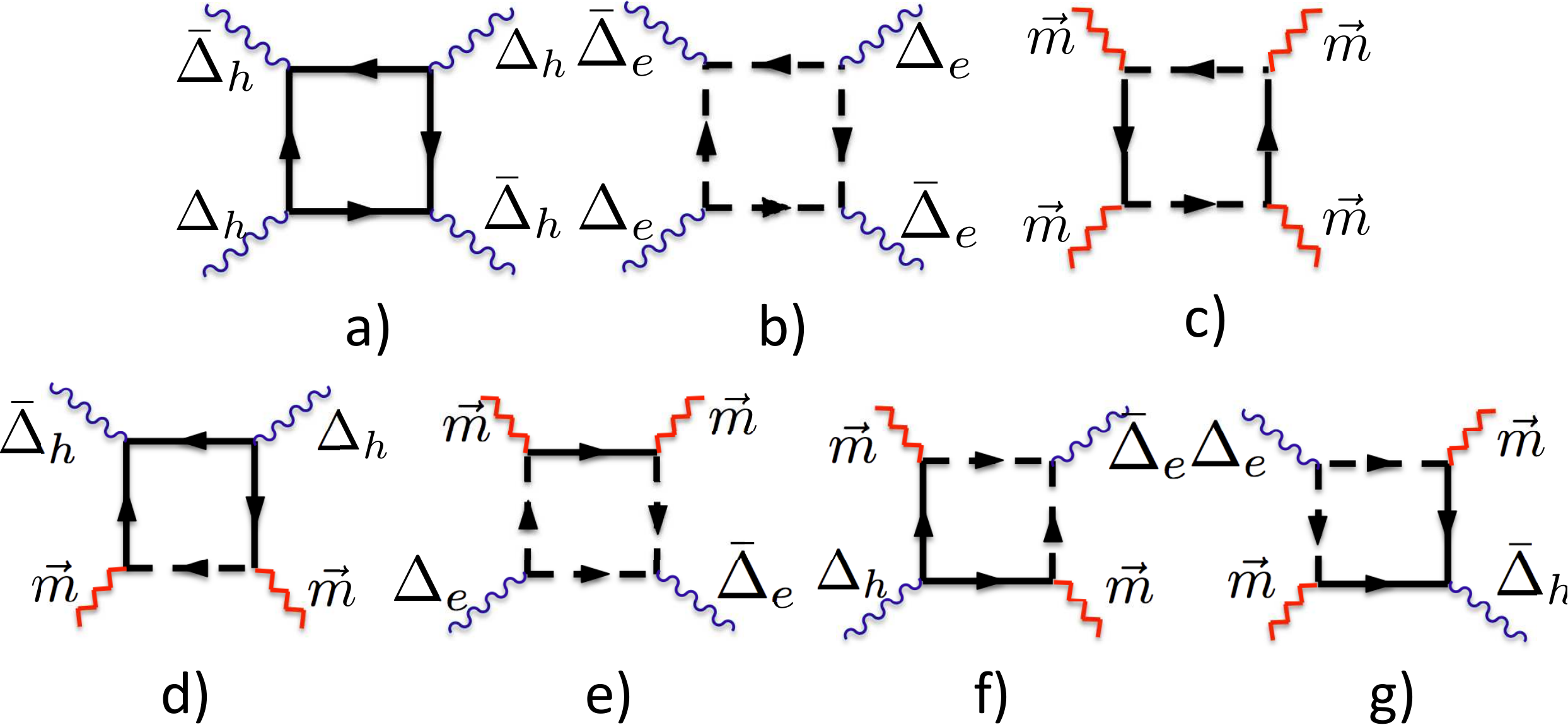}
\end{center}
\caption{(Color online) Interaction between fluctuations: wavy lines represent the superconducting fluctuators $L_{e/h}(q,\Omega, T)$,  broken lines stand for magnetic (SDW) fluctuator
$D_{m}(q,\Omega, T)$. Diagrams (a,b) define $A$ coefficient, (c) - $B$- coefficient, (d,e) - $C_1$ -  coefficient, (f,g) - $C_2$ - coefficient. Notations for solid and dashed lines are the same as in Fig.2}
\label{gl_2}
\end{figure}

  \kk{Although the situation in 1111 and 122 pnictides is described in the framework above, we consider  for the sake of generality also the opposite limit $u_3 \ll u_s$ and $u_s\Pi_s \sim 1$. }
In this case one deals with the three-mode coupling theory: two superconducting fluctuating modes $\Delta_e$ and $\Delta_h$ and one SDW $\vec m$ magnetic mode:
\begin{eqnarray}
{\cal L}^{(3)}=\int d\vec x \left\{\bar \Delta_e L^{-1}_e \Delta_e + \bar \Delta_h L^{-1}_h \Delta_h +
\vec m D^{-1}_m \vec m
+ A\left(|\Delta_e|^4+|\Delta_h|^4\right)+ B\vec m^4 +
C_1 \left(|\Delta_e|^2+|\Delta_h|^2\right)\vec m^2
+ C_2 \left(\bar\Delta_e\Delta_h+\bar\Delta_h\Delta_e\right)\vec m^2 \right\},
\label{l2}
\end{eqnarray}
with
\begin{eqnarray}
L^{-1}_e(q,\Omega,T)&=&L^{-1}_h(q,\Omega,T) =\nu\left[-\frac{i\Omega}{\gamma_{sc}} +\tau_{c} + c_\Delta q^2\right].
\end{eqnarray}
\kk{As in the static case, we neglect the difference between the contribution of electron and hole pockets in the fluctuation modes. }
In principle, the SDW and CDW modes are degenerate when $u_3 \Pi_1\ll 1$ and one needs to consider a four-mode  coupling theory. We however, assume that the SDW-CDW degeneracy is lifted out by
{\color{black} additional inter-band processes $u_2$  (see  Eq. (\ref{fullHamiltonian}) and further details in  Refs. \cite{CheEE08, MaChub10}), omitted through the derivation of the BS equations,} $T_{CDW} < T_{s}$  and restrict ourselves by the three-mode coupling theory.

The Lagrangian (\ref{l2}) at $C_2=0$ (Fig. \ref{gl_2} a-e) describes $U(1)\times U(1) \times SU(2)$ gauge theory where each $U(1)$ corresponds to the gauge symmetry of the superconducting (e- and h-) sectors and $SU(2)$ represents the magnetic (SDW) sector of the effective model. The most important observation is that magnetic fluctuations (Fig. \ref{gl_2} f-g) break $U(1)\times U(1) \to U(1)$ and mediate the cross-talk between two different superconducting (e-h) sectors.

\subsection*{Fluctuation corrections to $T_{c}$ and $T_{\rm s}$ }
In order to find the fluctuation correction to the superconducting transition temperature given by the mean-field analysis in the vicinity of the tetracritical point, we integrate out the {\color{black} remaining} slow magnetic
fluctuations in (\ref{l1}) {\color{black} and finally obtain the effective Lagrangian describing the superconducting system:}
\begin{eqnarray}
{\cal L}^{(2)}_\Delta&=& \int d\vec x \left\{\bar\Delta_- \tilde L^{-1}_{\pm} \Delta_- +
\tilde A|\Delta_-|^4\right\},
\end{eqnarray}
where
\begin{eqnarray}
\tilde L^{-1}_{\pm} = L^{-1}_{\pm} + K_1 +K_2,
\end{eqnarray}
is the superconducting fluctuator renormalized by magnetic fluctuations.
As a result, we find a fluctuation correction to $T_{\rm c}$:
\begin{equation}
T_{\rm c} =T_{\rm c}^0\cdot(1-\nu^{-1}(K_1+K_2)),
\end{equation}
where $T_{\rm c}^0$ is the transition temperature defined by the static solution of BS equations (mean-field theory) and
\begin{eqnarray}
K_{1} &=& T^2\sum_{nm}\int \frac{d\vec k d\vec q}{(2\pi)^{2d}} G^0_e(k,i\epsilon_n)G^0_e(k,i\epsilon_n)
G^0_e(-k,-i\epsilon_n)
 D_m(q,i\Omega_m) G^0_h(k-q,i\epsilon_n-i\Omega_m),\\
K_2 &=& - \frac{1}{2}T^2\sum_{nm}\int \frac{d\vec k d\vec q}{(2\pi)^{2d}} G^0_e(k,i\epsilon_n)G^0_e(-k,-i\epsilon_n) D_m(q,i\Omega_m)
 G^0_h(k-q,i\epsilon_n-i\Omega_m)G^0_h(-k+q,-i\epsilon_n+i\Omega_m).
\end{eqnarray}
{\color{black} Here $G^0_{e/h}(k,i\epsilon_n)=[i\epsilon_n - (\epsilon^{e/h}(k)-\mu)]^{-1}$ are bare Green's Functions for electron/hole bands respectively.}
We observe that there are two types of competing processes: the one with $+C_1$ leads to $K_1>0$ and results in suppression of $T_{\rm c}$ while the second one with $-C_2$ corresponds to $K_2<0$ and therefore
enhances $T_{\rm c}$.

\begin{figure}[t]
\begin{center}
\includegraphics[width=.7\columnwidth,angle=0]{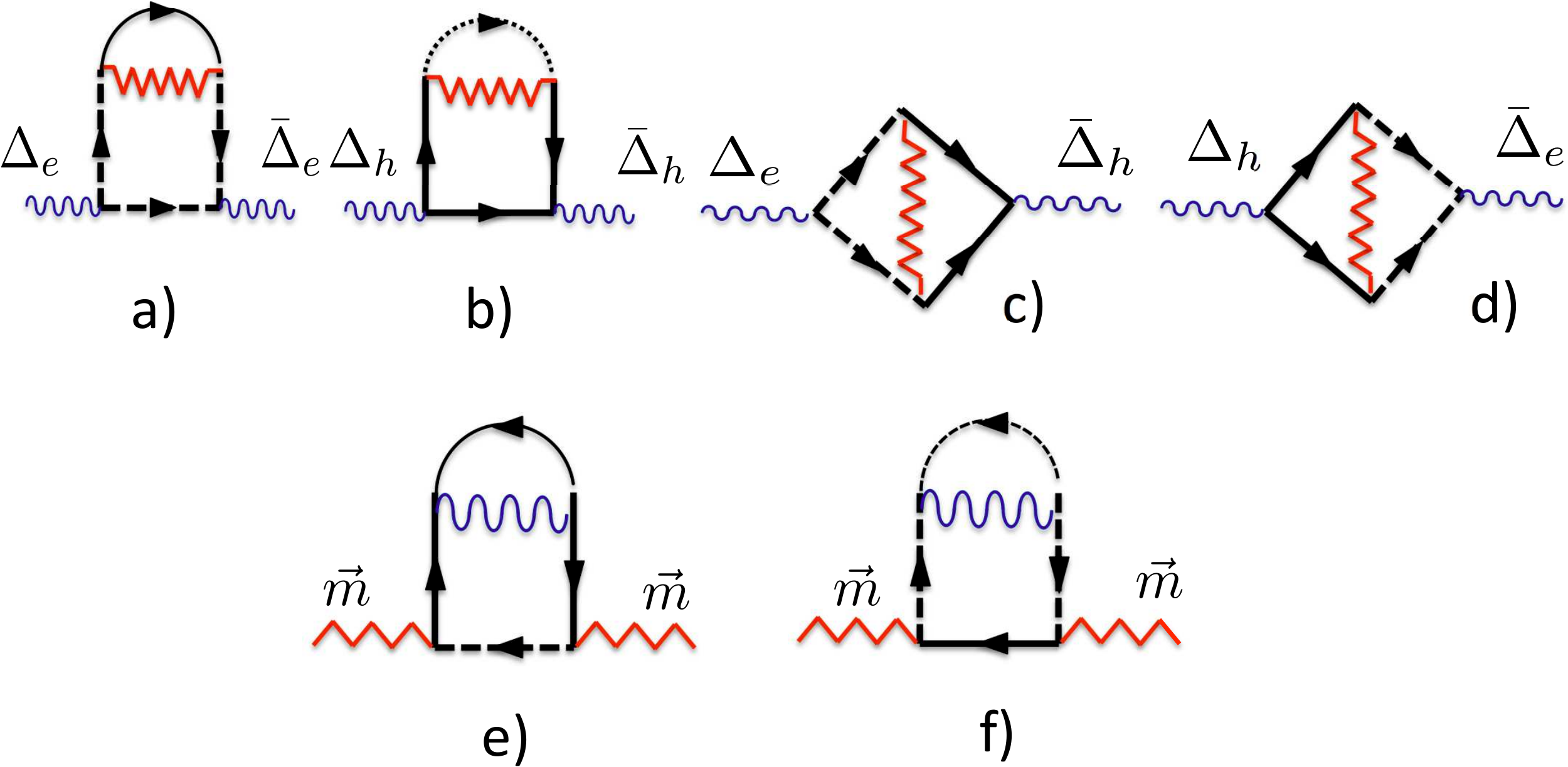}
\end{center}
\caption{(Color online) SDW fluctuation corrections to: e- and h- superconducting fluctuators (a,b), cross-coupled e-h terms (c,d). SC fluctuation corrections to the SDW fluctuators (e,f).}
\label{se_1}
\end{figure}

Similarly, for the second case described by the Lagrangian (\ref{l2}) characterized by magnetic fluctuation broken  $U(1)\times U(1)$ we get the following effective Lagrangian
\begin{eqnarray}
{\cal L}^{(3)}_{\Delta}&=&\int d\vec x\left\{
\left(
\begin{array}{cc}
\bar\Delta_e & \bar\Delta_h
\end{array}
\right)
\left(
\begin{array}{cc}
\tilde L^{-1}_e & K_2\\
K_2 & \tilde L^{-1}_h\\
\end{array}
\right)\left(
\begin{array}{c}
\Delta_e\\
\Delta_h\\
\end{array}
\right)
+ \tilde A\left(|\Delta_e|^4+|\Delta_h|^4\right)\right\},
\label{lag_s}
\end{eqnarray}
where the diagonal $\Delta_e - \Delta_e$ and $\Delta_h - \Delta_h$ inverse fluctuators
$\tilde L^{-1}_e +K_{1e}$  and $\tilde L^{-1}_h +K_{1h}$  are given by
the diagrams Fig. \ref{se_1} (a,b)  while the off-diagonal $\Delta_e - \Delta_h$ coupling $K_2$ is defined by the diagrams Fig. \ref{se_1} (c,d).
The corresponding equations for the $d$- dimensional cases ($d=2,3$) are given by:
\begin{eqnarray}
K_{1e} &=& T^2\sum_{nm}\int \frac{d\vec k d\vec q}{(2\pi)^{2d}} G^0_e(k,i\epsilon_n)G^0_e(k,i\epsilon_n)
G^0_e(-k,-i\epsilon_n)
D_m(q,i\Omega_m) G^0_h(k-q,i\epsilon_n-i\Omega_m),\\
K_{1h} &=& T^2\sum_{nm}\int \frac{d\vec k d\vec q}{(2\pi)^{2d}} G^0_h(k,i\epsilon_n)G^0_h(k,i\epsilon_n)
G^0_h(-k,-i\epsilon_n)
D_m(q,i\Omega_m) G^0_e(k-q,i\epsilon_n-i\Omega_m),\\
K_{2eh} &=& T^2\sum_{nm}\int \frac{d\vec k d\vec q}{(2\pi)^{2d}} G^0_e(k,i\epsilon_n)G^0_e(-k,-i\epsilon_n) D_m(q,i\Omega_m)
G^0_h(k-q,i\epsilon_n-i\Omega_m)G^0_h(-k+q,-i\epsilon_n+i\Omega_m).
\end{eqnarray}
While the diagrams Fig. \ref{se_1} (a,b) always reduce the effective temperature of the superconducting transition, the diagrams Fig. \ref{se_1} (c,d) lift the degeneracy between e-h transition temperatures.
As a result, assuming that $K_{1e}=K_{1h}=K_{1}$, $K_{2eh}=K_2$ we get the fluctuation correction to the critical temperature $T_c$:
\begin{equation}
T_c =T_c^0\cdot(1+\nu^{-1}(|K_2|-|K_1|)).
\label{tcs1}
\end{equation}

To construct the effective field theory for the influence of superconducting fluctuations on the SDW dynamics, we integrate out the slow superconducting fluctuations in (\ref{l2})
{\color{black} and finally obtain the effective Lagrangian for paramagnetic SDW fluctuations}:
\begin{eqnarray}
{\cal L}^{\vec m}_{eff}=\int d \vec x \left\{\vec m \tilde D_{m}^{-1} \vec m + \tilde B \vec m^4\right\}
\label{lag_m}
\end{eqnarray}
where
for two-mode coupling theory $\tilde D_{m}^{-1}=D_{m}^{-1}+\tilde K_{1}+\tilde K_{2}$ with
\begin{eqnarray}
\tilde K_{1} &=& T^2\sum_{nm}\int \frac{d\vec k d\vec q}{(2\pi)^{2d}} G^0_e(k,i\epsilon_n)G^0_e(k,i\epsilon_n)
G^0_h(k,i\epsilon_n)
L_\pm(q,i\Omega_m) G^0_e(q-k,i\Omega_m-i\epsilon_n),\\
\tilde K_{2} &=& - \frac{1}{2}T^2\sum_{nm}\int \frac{d\vec k d\vec q}{(2\pi)^{2d}} G^0_e(k,i\epsilon_n)
G^0_e(-k,-i\epsilon_n) L_\pm(q,i\Omega_m)
G^0_h(k-q,i\epsilon_n-i\Omega_m)G^0_h(-k+q,-i\epsilon_n+i\Omega_m),
\end{eqnarray}
and
\begin{equation}\label{sdwred1}
T_{s} =T_{s}^0\cdot(1-\nu^{-1}(\tilde K_1 + \tilde K_2))
\end{equation}
while for three-mode coupling theory
 $\tilde D_{m}^{-1}=D_{m}^{-1}+\tilde K_{1e}+\tilde K_{1h}$ with
\begin{eqnarray}
\tilde K_{1e} &=& T^2\sum_{nm}\int \frac{d\vec k d\vec q}{(2\pi)^{2d}} G^0_e(k,i\epsilon_n)G^0_e(k,i\epsilon_n)
G^0_h(k,i\epsilon_n)
L_e(q,i\Omega_m) G^0_e(q-k,i\Omega_m-i\epsilon_n),\\
\tilde K_{1h} &=& T^2\sum_{nm}\int \frac{d\vec k d\vec q}{(2\pi)^{2d}} G^0_h(k,i\epsilon_n)G^0_h(k,i\epsilon_n)
G^0_e(k,i\epsilon_n)
L_h(q,i\Omega_m) G^0_h(q-k,i\Omega_m-i\epsilon_n).
\end{eqnarray}
The diagrams defining the fluctuation corrections to $T_{\rm s}$ are shown on Fig. \ref{se_1} (e,f).

Computing the correcting terms for the three-mode coupling theory we obtain that the fluctuation correction to the SDW transition temperature always leads to its reduction:
\begin{equation}\label{sdwred1a}
T_{\rm s} =T_{\rm s}^0\cdot(1-2\nu^{-1}|\tilde K_1|).
\end{equation}
Exactly at the tetracritical point $T_{\rm c}^0=T_{\rm s}^0=T_0$, {\color{black} the concentration $c=c_0$}, $D_{m}^{-1}=L_{e}^{-1}=
L_{h}^{-1}$. Due to the assumed particle-hole symmetry we get $\tilde K_1 = K_1$. Finally,
combining (\ref{tcs1}) and (\ref{sdwred1a}) we get
\begin{equation}\label{sdwred2}
T_{\rm c} - T_{\rm s} =T_0\cdot \nu^{-1}\cdot(|K_1| + |K_2|)>0,
\end{equation}
and therefore the new tetracritical point corresponds to lower values of the concentration $c<c_0$ (see Fig. \ref{fig_2}).

{\color{black} Let us illustrate, as an example, the calculation of the fluctuation corrections diagrams $K_{1,2}$ containing the superconducting fluctuator $L_{e/h}(q,\Omega)$ (the evaluation of the diagrams containing the SDW fluctuator can be performed in a similar fashion).
The evaluation of the Matsubara sums and integrals can be done in several steps:\\
i) First, the main contribution to the sum over $\Omega_m$ is given by the $m=0$ term. Therefore we can replace
\begin{eqnarray}
T\sum_m L_e(q,i\Omega_m) G^0_e(q-k,i\Omega_m-i\epsilon_n)\to T_0 L_e(q,0) G^0_e(q-k,-i\epsilon_n),  \nonumber\\
T\sum_m L_h(q,i\Omega_m) G^0_h(q-k,i\Omega_m-i\epsilon_n)\to T_0 L_h(q,0) G^0_h(q-k,-i\epsilon_n).
\end{eqnarray}
ii) Second, we notice that the integral over ${\bf q}$ is determined by the small $q$: $q \xi_{s}\sim \sqrt{\tau_c}\ll 1$ where
$\xi_{s}$ is the superconducting coherence length,  $\xi_{s,d}^2=(7\zeta(3) v_F^2)/(16 d \pi^2 T_0^2)$ for d-spatial dimensions $d=2, 3$. We can therefore neglect the $q$-dependence in the $e/h$ Green's functions:
$$
G^0_{e/h}({\bf q-k},-i\epsilon_n)\to G^0_{e/h}({\bf -k},-i\epsilon_n).
$$
iii) Next, the sum over
fermionic Matsubara frequency $\epsilon_n$ and summation over ${\bf k}$ in $K_1$ and $K_2$ is performed in the same way as explained below in the Section "Ginzburg-Landau approach" (see Section "Methods"). As a result we obtain:
\begin{equation}
T\sum_n \int \frac{d^d k}{(2\pi)^d} [G^{(0)}(-k,-i\epsilon_n)]\cdot
[G^{(0)}(k,i\epsilon_n)]^3 =
T\sum_n \int \frac{d^d k}{(2\pi)^d} [G^{(0)}(-k,-i\epsilon_n)]^2\cdot
[G^{(0)}(k,i\epsilon_n)]^2\to \frac{7\zeta(3)\nu}{16\pi^2 T_0^2}
\end{equation}
Notice, that after these approximations there is no difference between $K_1$ and $K_2$.\\
iv) The remaining integral over momentum ${\bf q}$ is performed using the {\it static} Ornstein-Zernike correlator $L_{e/h}(q,0)$ (see \cite{Varla})
\begin{equation}
\sum_q L_{e/h}(q,0)\to V_d \int_0^{\xi_{s,d}^{-1}}\frac{q^{d-1} dq}{(2\pi)^d} \frac{\nu^{-1}}{c_\Delta q^2+\tau_c},
\label{oz1}
\end{equation}
here $V_d$ denotes the unit volume. We see that for $d=3$ the integral (\ref{oz1}) is convergent,
\begin{equation}
\nu^{-1} K_{1,2} \sim \frac{1}{\nu T_0 \xi_{s,3}^3} \sim \sqrt{Gi^{(3)}}.
\end{equation}
Here we introduced the Ginzburg number $Gi^{(d)}$ defined as the range of reduced  temperatures $\delta T/T_c$ where the fluctuation corrections to the specific heat are comparable to its jump at $T_c$. For $d=3$ the Ginzburg number  $ Gi^{(3)} \sim (T_0/\varepsilon_F)^4$.

For $d=2$ the integral (\ref{oz1}) diverges logarithmically:
\begin{equation}
\nu^{-1} K_{1,2} \sim \frac{1}{\nu T_0 \xi_{s,2}^2}\ln \left(c_\Delta\xi_{s,2}^{-2}/\tau_c\right) \sim  Gi^{(2)} \ln \left(c_\Delta\xi_{s,2}^{-2}/\tau_c\right).
\label{oz2}
\end{equation}
v) Finally we replace $\tau_{c}$ in  (\ref{oz2}) by its low bound - the Ginzburg number (recall, that by its definition $Gi^{(d)}\ll \tau_c\ll 1$) and substitute $Gi^{(2)}\sim T_0/\varepsilon_F$  (see \cite{Varla} where similar calculations have been performed for a single $U(1)$-mode theory by means of Renormalization Group technique).

Performing similar calculations for the renormalization of the SDW critical temperature $T_s$ we obtain:
\begin{eqnarray}
2D &:& \frac{T_{\rm c}-T_{\rm s}}{T_0} \sim - Gi^{(2)}\ln\left( Gi^{(2)}\right)\sim\frac{T_0}{\epsilon_F}\ln\left(\frac{\epsilon_F}{T_0}\right),\nonumber\\
3D &:& \frac{T_{\rm c}-T_{\rm s}}{T_0} \sim  \sqrt{Gi^{(3)}} \sim \left(\frac{T_0}{\epsilon_F}\right)^2.
\label{eq_35}
\end{eqnarray}
We emphasize that the constants $K_1$ and $K_2$ characterizing the fluctuational shift of the critical temperature are found to depend on the Ginzburg number $Gi^{(d)}$
and the dimensionality of the system $d$ only.}

\kk{Eqs. (\ref{eq_35}) represent the central result of the paper: fluctuation corrections are responsible for the competition between the spin density wave and superconducting critical modes in the vicinity of the tetracritical point. We have obtained this results by calculating the fluctuation corrections to the GL functional, but the same result may be obtained in the framework of renormalization group approach.}
Since the $U(1)\times U(1)$ symmetry is {\it a priori} broken in the superconducting sector of the model, the influence of the magnetic fluctuations onto the superconducting transition is two-fold: first, the intra-band corrections tempt to reduce the transition temperature; second -  the inter-band corrections do exactly the opposite - split the two transition temperatures and effectively facilitate the superconducting transition. On the other hand, the contribution of superconducting fluctuations in the $SU(2)$ magnetic sector of the model does result only in a suppression of the SDW transition. The contribution of "eigen" fluctuations, namely the superconducting-superconducting and magnetic-magnetic is alike and therefore it is dropped off from the difference of the critical temperatures.
The asymmetric character of the fluctuation corrections results in the shift of tetracritical point towards lower values of the carriers concentration. For the "dirty" limit of the multi-mode coupling theory the Ginzburg numbers should be updated accordingly \cite{Varla}.

 \de{The strong renormalizations of $T_c$ and $T_s$  are expected  in iron based superconductors, which are rather two-dimensional and many of them have rather small Fermi energy. In these materials the value of the Ginzburg number may be quite large.
For instance, the characteristic value of the Fermi energy in 1111 and 11 compounds is 100 meV \cite{Charnukha2015,Lubashevsky2012,Ozaki2014}, while the superconducting critical temperature is about 20-50 K. With this numbers the one finds large Ginzburg number $Gi^{(2)} \sim T_c/\epsilon_F \sim 0.02-0.05 $ and  considerable renormalizations of the critical temperatures$(T_c -T_s)/T_c \sim 0.1$. }

\section*{Discussion}

 We have developed a high-temperature approach to the problem of interplay between magnetic and superconducting ordering in multi-band systems. Both static and dynamical (fluctuation related) contribution to the mode-mode coupling are discussed. It is shown that the fluctuation corrections in the vicinity of the tetracritical point reduce the magnetic transition in accordance with Eqs. (\ref{sdwred1}), (\ref{sdwred2}). On the superconductor side of the phase diagram the situation is different. The intraband and interband contributions of spin fluctuations to the Ginzburg-Landau functional nearly compensate each other.

The approach that we have introduced is more general than for instance a mean-field description of competing phases \cite{HiKoMa11,Chubuk11} as it accounts for dynamical fluctuations and goes beyond the static scaling paradigm. Apart from this, the framework presented here has the advantage that it can be generalized to other multi-mode regimes in a straightforward manner, which can include e.g. the presence of a nematic instability, the competition between $s_\pm$ and $s_{++}$ and/or singlet/triplet pairings. {\color{black} Besides, it is highly appealing to discuss in a framework of this approach  the influence of conventional magnetic defects (transition metal ions substituting for Fe) and weak magnetic defects produced by Zn impurities and As-vacancies on the superconducting transition temperature. This work is now in progress.}

\section*{Methods}
\label{sec:methods}
\subsection*{Integral Bethe-Salpeter equations}
\label{sec:apa}
The integral Bethe-Salpeter equations in $d=2,3$ dimensions take the form
\begin{eqnarray}
\Gamma_{11}^{\sigma\sigma}(p_1,p_2,p_3,p_4)&=&
\mathfrak{J}_{10}^{\sigma\sigma}(p_1,p_2,p_3,p_4)+T\sum_n\int\frac{d^d k}{(2\pi)^d}
\mathfrak{J}_{10}^{\sigma\sigma}(k,p_2,p_3,k+p_2-p_3)
\Pi_1(k,p_2-p_3)\Gamma_{11}^{\sigma\sigma}(p_1,k+p_2-p_3,k,p_4)\nonumber\\
&+&T\sum_n\int\frac{d^d k}{(2\pi)^d} \mathfrak{J}_{30}^{\sigma\bar\sigma}(p_2,k+p_2-p_3,p_3,k)
 \Pi_1(k,p_2-p_3) \Gamma_{32}^{\sigma\bar\sigma}(p_1,k,p_4,k-p_2+p_3),\nn\\
 \Gamma_{12}^{\sigma\bar\sigma}(p_1,p_2,p_3,p_4)&=&T\sum_n\int\frac{d^d k}{(2\pi)^d}
 \mathfrak{J}_{30}^{\sigma\bar\sigma}(p_2,k-p_2+p_3,p_3,k)
 \Pi_1(k,p_2-p_3) \Gamma_{31}^{\sigma\sigma}(k,p_1,k-p_2+p_3,p_4), \nn \\
 \Gamma_{31}^{\sigma\sigma}(p_1,p_2,p_3,p_4)&=&\mathfrak{J}_{30}^{\sigma\sigma}(p_1,p_2,p_3,p_4)
 +T\sum_n\int\frac{d^d k}{(2\pi)^d} \mathfrak{J}_{30}^{\sigma\bar\sigma}(p_1,k,p_3,k+p_1-p_3)
 \Pi_1(k,p_1-p_3) \Gamma_{12}^{\sigma\bar\sigma}(p_2,k+p_1-p_3,p_4,k)\nonumber\\
  &+&T\sum_n\int\frac{d^d k}{(2\pi)^d} \mathfrak{J}_{10}^{\sigma\sigma}(p_1,k-p_1+p_3,k,p_3)
 \Pi_1(k,p_1-p_3) \Gamma_{31}^{\sigma\bar\sigma}(p_2,k,p_4,k-p_1+p_3), \nn \\
 \Gamma_{32}^{\sigma\bar\sigma}(p_1,p_2,p_3,p_4)&=&\mathfrak{J}_{30}^{\sigma\bar\sigma}(p_1,p_2,p_3,p_4)
 +T\sum_n\int\frac{d^d k}{(2\pi)^d} \mathfrak{J}_{30}^{\sigma\bar\sigma}(p_1,k,p_3,k+p_1-p_3)
 \Pi_1(k,p_1-p_3) \Gamma_{11}^{\bar\sigma\bar\sigma}(p_2,k+p_1-p_3,p_4,k)\nonumber\\
 &+&T\sum_n\int\frac{d ^3k}{(2\pi)^d} \mathfrak{J}_{10}^{\sigma\sigma}(p_1,k-p_1+p_3,k,p_3)
 \Pi_1(k,p_1-p_3) \Gamma_{32}^{\sigma\bar\sigma}(k,p_2,k-p_1+p_3,p_4)\nonumber\\
 &+&T\sum_n\int\frac{d^d k}{(2\pi)^d} \mathfrak{J}_{30}^{\sigma\bar\sigma}(p_1,p_2,k,-k+p_1+p_2)
 \Pi_s(k,p_1+p_2) \Gamma_{4}^{\sigma\bar\sigma}(k,p_1+p_2-k,p_3,p_4)\nonumber \\
 &+&T\sum_n\int\frac{d^d k}{(2\pi)^d} \mathfrak{J}_{s0}^{\sigma\bar\sigma}(p_1,p_2,k,-k+p_1+p_2)
 \Pi_s(k,p_1+p_2) \Gamma_{32}^{\sigma\bar\sigma}(k,p_1+p_2-k,p_3,p_4), \label{eq1.app.bethe-salpeter}\\
 \Gamma_{4}^{\sigma\bar\sigma}(p_1,p_2,p_3,p_4)&=&\mathfrak{J}_{s0}^{\sigma\bar\sigma}(p_1,p_2,p_3,p_4)
 +T\sum_n\int\frac{d^d k}{(2\pi)^d} \mathfrak{J}_{s0}^{\sigma\bar\sigma}(p_1,p_2,k,-k+p_1+p_2)
 \Pi_s(k,p_1+p_2) \Gamma_{4}^{\sigma\bar\sigma}(k,p_1+p_2-k,p_3,p_4)\nonumber \\
 &+&T\sum_n\int\frac{d^d k}{(2\pi)^d} \mathfrak{J}_{30}^{\sigma\bar\sigma}(p_1,p_2,k,-k+p_1+p_2)
 \Pi_s(k,p_1+p_2) \Gamma_{32}^{\sigma\bar\sigma}(k,p_1+p_2-k,p_3,p_4),\nn \\
 \Gamma_{5}^{\sigma\bar\sigma}(p_1,p_2,p_3,p_4)&=&\mathfrak{J}_{s0}^{\sigma\bar\sigma}(p_1,p_2,p_3,p_4)
 +T\sum_n\int\frac{d^d k}{(2\pi)^d} \mathfrak{J}_{s0}^{\sigma\bar\sigma}(k,p_1+p_2-k,p_3,p_4)
 \Pi_s(k,p_1+p_2) \Gamma_{5}^{\sigma\bar\sigma}(p_1,p_2,k,p_1+p_2-k)\nonumber\\
 &+&T\sum_n\int\frac{d^d k}{(2\pi)^d} \mathfrak{J}_{30}^{\sigma\bar\sigma}(k,p_1+p_2-k,p_3,p_4)
 \Pi_s(k,p_1+p_2) \Gamma_{32}^{\sigma\bar\sigma}(p_1,p_2,k,p_1+p_2-k), \nn
\end{eqnarray}
where we used the short-hand notations
\begin{eqnarray*}
\Pi_1(k,q)&=& G_e(k) G_h(k+q)=G_h(k) G_e(k+q),\;\;\;\;\;\;\;\;\;\;\;\;\;
\Pi_s(k,q)= G_e(k) G_e(q-k) =  G_h(k) G_h(q-k),\;\;\;\;\;\;\;\;\;\;\;\;
\end{eqnarray*}
$p_i=(\vec p_i, i\omega_{n_{i}})$, $k=(\vec k, i\epsilon_{n})$, $q=(\vec q, i\Omega_{m})$ with fermionic Matsubara frequencies $\omega_n=\epsilon_n=2\pi T(n+1/2)$ and bosonic Matsubara frequency
$\Omega_m=2\pi m T$  and the irreducible vertices $\mathfrak{J}_{j0}$ include all irreducible diagrams in the channel $j$. The spin, momentum and
energy are conserved in each vertex: $p_4=p_1+p_2-p_3$.
The simplified matrix BS equations are obtained by a replacement
\begin{eqnarray*}
\mathfrak{J}_{10}\to u_1,\;\;\;\;\;
\mathfrak{J}_{30}\to u_3,\;\;\;\;\;
\mathfrak{J}_{s0}\to u_s,
\end{eqnarray*}

Due to momentum conservation there are only three independent momenta. It worth to introduce new variables $p = p_1 + p_2=p_3+p_4$,
$q = p_3 - p_2$ and $t = p_3 - p_1$. In the new variables $\tilde\Gamma_{11}(p,q,t) = \Gamma_{11}(p_1,p_2,p_3,p_4)$ the first integral in Eq.(\ref{eq1.app.bethe-salpeter}) has the form:

\ba\Gamma_{11}^{\sigma\sigma}(p,q,t)
&=& u_1+u_1 T\sum_n\int\frac{d^dk}{(2\pi)^d} \tilde
\Pi_1(k,q)\tilde\Gamma_{11}^{\sigma\sigma}(k+(p- q-t)/2,q,k-(p+q-t)/2)\nn
\\&+&u_3 T\sum_n\int\frac{d^3k}{(2\pi)^3} \tilde
\Pi_1(k,q)\Gamma_{32}^{\sigma\bar\sigma}(k+(q+p-t)/2,(p-t-q)/2-k,-q),
\nn \\
 \Gamma_{32}^{\sigma\bar\sigma}(p,q,t)&=& u_3
 + u_3 T\sum_n\int\frac{d^d k}{(2\pi)^d}
 \Pi_1(k,t) \Gamma_{11}^{\bar\sigma\bar\sigma}(1/2(p-t-q) + k, -k + 1/2(p+t-q),-t) \nonumber\\
 &+&u_1 T\sum_n\int\frac{d ^3k}{(2\pi)^d}
 \Pi_1(k,t) \Gamma_{32}^{\sigma\bar\sigma}(1/2(p+t-q)+k,k+1/2(t+q-p),t)\nonumber\\
 &+&u_3 T\sum_n\int\frac{d^d k}{(2\pi)^d}
 \Pi_s(k,p) \Gamma_{4}^{\sigma\bar\sigma}(p,1/2(q+t-p)+k,1/2(q+t+p)-k )\label{eq.app.bethe-salpeter} \\
 &+&u_s T\sum_n\int\frac{d^d k}{(2\pi)^d}
 \Pi_s(k,p) \Gamma_{32}^{\sigma\bar\sigma} (p,1/2(q+t-p)+k,1/2(q+t+p)-k),\nn \\
 \Gamma_{4}^{\sigma\bar\sigma}(p,q,t)&=& u_s
 +u_s T\sum_n\int\frac{d^d k}{(2\pi)^d}
 \Pi_s(k,p) \Gamma_{4}^{\sigma\bar\sigma}(p,1/2(q+t-p)+k,1/2(q+t+p)-k))\nonumber \\
 &+& u_3 T\sum_n\int\frac{d^d k}{(2\pi)^d}
 \Pi_s(k,p) \Gamma_{32}^{\sigma\bar\sigma}(p,1/2(q+t-p)+k,1/2(q+t+p)-k)). \nn
\ea

Considering the first equation one sees that the dependence  $\Gamma_{11}^{\sigma\sigma}(p,q,t)$ on the first $p$ and the last $t$ variables is weak due to internal integrations over $\bar k$ and
can be substituted with $log$ accuracy by the characteristic transfer momenta $\bar k = const$.
It leads $\Gamma_{11}^{\sigma\sigma}(p,q,t) \to \Gamma_{11}^{\sigma\sigma}(\bar k, q,  \bar k)$.
Similar  one finds $\Gamma_{4}^{\sigma\bar\sigma}(p,q,t) \approx \Gamma_{4}^{\sigma\sigma}(p,\bar k,\bar k)$. {\color{black} However},   $\Gamma_{32}^{\sigma\bar\sigma}(p,q,t) \approx \Gamma_{32}^{\sigma\bar\sigma}(p,\bar k,t)$
vertex depends on two variables due to  couplings to the {\color{black} Cooper} and the SDW channels.
With this substitution and  using the bare Green's functions instead of dressed GF in the polarization loops, {\color{black} we obtain}:
\begin{eqnarray*}
\Pi_1(q) &\to& T\sum_n \int\frac{d^d k}{(2\pi)^d} \Pi_1(k,q),\\
\Pi_s(q) &\to& T\sum_n \int\frac{d^d k}{(2\pi)^d}  \Pi_s(k,q),
\end{eqnarray*}

the Eq.(\ref{eq.app.bethe-salpeter}) get a simplified form:
\ba
 \Gamma_{11}(\bar k,q,\bar k) &\approx& u_1 + (u_1   \Gamma_{11}(\bar k,q,\bar k) + u_3    \Gamma_{32}(\bar k,\bar k,q))\Pi_1 (q), \nn \\
 \Gamma_{32}(p,\bar k,q) &\approx& u_3 + (u_3  \Gamma_{11}(\bar k,q,\bar k) + u_1    \Gamma_{32}(\bar k,\bar k,q))\Pi_1 (q)  
+ (u_3    \Gamma_{4}(p,\bar k,\bar k) + u_4   \Gamma_{32}(p,\bar k,\bar k))\Pi_s (p), \\
 \Gamma_{4}(p,\bar k,\bar k) &\approx& u_4 + (u_4    \Gamma_{4}(p,\bar k,\bar k) + u_3    \Gamma_{32}(\bar k,\bar k,p))\Pi_s (p). \nn
\ea
In the case of $T\gg \bar k\cdot v_F$, SDW and SC channels are strongly coupled  and  have the same pole structure.
With logarithmic accuracy for $T\ll \bar k \cdot v_F \sim \epsilon_F$ the two channels separate from each other.

Now let us consider the case of finite temperature away from tetracritical point $\Pi_1(0)\gg \Pi_s(0)$.
The system of equations with log accuracy reduces to the system:
\ba
 \Gamma_{11}(\bar k,q,\bar k) &\approx& u_1 + (u_1   \Gamma_{11}(\bar k,q,\bar k) + u_3    \Gamma_{32}(\bar k,\bar k,q))\Pi_1 (q),\\
 \Gamma_{32}(\bar k,\bar k,t) &\approx& u_3 + (u_3  \Gamma_{11}(\bar k,q,\bar k) + u_1    \Gamma_{32}(\bar k,\bar k,q))\Pi_1 (q),
\ea
which can be easily solved
\ba
 \Gamma_{11}(\bar k,q,\bar k) \pm  \Gamma_{32}(\bar k,\bar k,q) &=& \frac{u_1 \pm u_3}{1- (u_1  \pm u_3)\Pi_1 (q)}.
\ea
Here the sign "$+$"  corresponds to SDW channel, while the the sign "$-$" to CDW channel. {\color{black} The} SDW magnetic instability develops when $1- (u_1  + u_3)\Pi_1 (q) = 0$.  Close to the instability point it is naturally to expand near the SDW vector ${\bf Q}$.
Then $1- (u_1  + u_3)\Pi_1 ({\bf Q+q}, i\Omega_n\to \Omega+i0^{+})  \propto \tau + c_m  q^2 -i\Omega/\gamma_{sdw}$ (see also Section "Ginzburg-Landau approach").
It leads
\ba
 \Gamma_{11}(\bar k,{\bf Q+q},\bar k) \sim  \Gamma_{32}(\bar k,\bar k,{\bf Q+q}) \to \Gamma_{sdw}({\bf Q+q},i\Omega_n
\to \Omega+i0^{+}) = \frac{1}{2}  \frac{u_1\pm u_3}{\tau_{s} + c_m  q^2 -i \Omega/\gamma_{sdw}}.
\ea

Similar to above we consider the  superconducting channel away from the tetracritical point: \be
 \Gamma_{4}(q,\bar k,\bar k) \pm  \Gamma_{32}(q,\bar k,\bar k) = \frac{u_s \pm u_3}{1- (u_s  \pm u_3)\Pi_s(q)},\\
\ee
{\color{black} where $+$ corresponds to  $s_{\pm}$ superconductivity, while $-$ to $s_{++}$}.
Close to the transition point for small  $q$ :
\ba
- \Gamma_{4}(q,\bar k,\bar k) \sim  \Gamma_{32}(q,\bar k,\bar k) \to   \Gamma_{32}(q,i\Omega_n\to \Omega+i0^{+})=
\frac{1}{2}\frac{u_s\pm u_3}{\tau_{s_\pm} +  c_\Delta q^2 -i\Omega/\gamma_{sc_\pm}}.
\ea

\subsection*{Effective action for the multi-mode theory: beyond the Bethe-Salpeter approach}
\label{sec:apb}
We start with the Hamiltonian (\ref{fullHamiltonian}). Since superconducting and magnetic sectors are block-diagonal, let's begin with description of the Gaussian action for two-band superconductor away from the magnetic phase:
\begin{eqnarray}
{\cal L}_{eh}(\tau)= \sum_\alpha\int d\vec x \left\{\bar f_\alpha \partial_\tau f_\alpha
+\bar c_\alpha \partial_\tau c_\alpha \right\} - H,\;\;\;\;\;\;\; S=\int_0^{1/T} d\tau {\cal L}(\tau).
\end{eqnarray}
As a first step we introduce fluctuating fields for description of the  e- ($f$) and h- ($c$) superconductivity
\begin{eqnarray}
\int {\cal D}\left[\bar\Delta_e \Delta_e \bar\Delta_h \Delta_h\right]
\delta\left(\Delta_e -f_\uparrow f_\downarrow\right)
\delta\left(\bar \Delta_e -\bar f_\downarrow \bar f_\uparrow\right)
\delta\left(\Delta_h -c_\uparrow c_\downarrow\right)
\delta\left(\bar \Delta_h - \bar c_\downarrow \bar c_\uparrow\right).
\end{eqnarray}
The second step is to use an integral representation for the functional $\delta$- functions
\begin{eqnarray}
\delta\left(\Delta_e -f_\uparrow f_\downarrow\right)&=&\int {\cal D}[\bar \eta_e] \exp\left(i\bar\eta_e\left[\Delta_e-f_\uparrow f_\downarrow\right] \right),\;\;\;\;\;
\delta\left(\bar \Delta_e -\bar f_\downarrow \bar f_\uparrow\right)=\int {\cal D}[\eta_e] \exp\left(-i\eta_e\left[\bar \Delta_e -\bar f_\downarrow \bar f_\uparrow\right] \right),\nonumber\\
\delta\left(\Delta_h -c_\uparrow c_\downarrow\right)&=&\int {\cal D}[\bar \eta_h] \exp\left(i\bar\eta_h\left[\Delta_h-c_\uparrow c_\downarrow\right] \right),\;\;\;\;\;
\delta\left(\bar \Delta_h - \bar c_\downarrow \bar c_\uparrow\right)=\int {\cal D}[\eta_h] \exp\left(-i\eta_h\left[\bar \Delta_h -\bar c_\downarrow \bar c_\uparrow\right] \right),
\end{eqnarray}
and the Lagrangian takes form
\begin{eqnarray}
{\cal L}_{fc\Delta\eta} = \int d\vec x\left\{
-\left(
\begin{array}{cc}
\bar\Delta_e & \bar\Delta_h
\end{array}
\right)
\left(
\begin{array}{cc}
u_s & u_3\\
u_3 & u_s\\
\end{array}
\right)\left(
\begin{array}{c}
\Delta_e\\
\Delta_h\\
\end{array}
\right)+
\bar\Psi^T \hat M \Psi
+ i\bar \eta \Delta + c.c.\right\},
\nonumber
\end{eqnarray}
where
\begin{eqnarray}
\bar \Psi^T=
\left(
\begin{array}{cccc}
\bar f_\uparrow & f_\downarrow &\bar c_\uparrow & c_\downarrow
\end{array}
\right)^T
\end{eqnarray}
is a Nambu 2-spinor and
\begin{eqnarray}
\hat M (\eta) =\left(
\begin{array}{cccc}
\hat G_e^{-1} & i\bar \eta_e & 0 & 0  \\
-i\eta_e & \breve{G}_e^{-1}& 0& 0 \\
0 & 0 & \hat G_h^{-1} & i\bar \eta_h  \\
0 & 0& -i\bar\eta_h  & \breve{G}_h^{-1}
\end{array}
\right),
\end{eqnarray}
and {\color{black} $\hat G_{e/h}({\bf k},i\epsilon_n) = [i\epsilon_n -(\epsilon^{e/h}_k-\mu)]^{-1}$ and $\breve G_{e/h}({\bf k},i\epsilon_n) = [-i\epsilon_n -(\epsilon^{e/h}_k-\mu)]^{-1}$.}

As a next step we compute Gaussian integral over Grassmann fields $f$ and $c$. As a result,
\begin{eqnarray}
{\cal L}_{\Delta\eta} &=& \int d\vec x\left\{
-\left(
\begin{array}{cc}
\bar\Delta_e & \bar\Delta_h
\end{array}
\right)
\left(
\begin{array}{cc}
u_s & u_3\\
u_3 & u_s\\
\end{array}
\right)\left(
\begin{array}{c}
\Delta_e\\
\Delta_h\\
\end{array}
\right) + i\bar \eta \Delta + c.c. \right\} + {\rm Tr Log} \left[\hat M (\eta) \hat M^{-1}(0)\right].
\end{eqnarray}
Expanding ${\rm Tr Log}\left[\hat M (\eta) \hat M^{-1}(0)\right]$
in terms of fermionic loops one gets:
\begin{eqnarray}
{\rm Tr Log} \left[\hat M (\eta) \hat M^{-1}(0)\right] = -{\rm Tr}\sum_{k=1}^{\infty}
\frac{\left[\hat G_0(\hat M(\eta)- \hat G_0^{-1})\right]^{2k}}{2k},
\end{eqnarray}
where expansion contains even number of the Green's functions {\color{black} $G_0=diag[\hat G_e,\hat G_h,\breve G_e,\breve G_h]$} and for minimal mode-mode coupling theory it is enough to retain quadratic and quartic terms.
Finally, computing the path integral over $\eta$'s by the saddle point:
\begin{eqnarray}
\frac{\partial {\cal L}_{\Delta\eta}}{\partial \bar \eta_e} &=& i\Delta_e + \frac{\partial}{\partial\bar \eta_e} {\rm Tr Log} \left[\hat M (\eta) \hat M^{-1}(0)\right]=0,\nonumber\\
\frac{\partial {\cal L}_{\Delta\eta}}{\partial \bar \eta_h} &=& i\Delta_h + \frac{\partial}{\partial\bar \eta_h} {\rm Tr Log} \left[\hat M (\eta) \hat M^{-1}(0)\right]=0,
\end{eqnarray}
we arrive at the effective action describing fluctuating Cooper pairs:
\begin{eqnarray}
{\cal L}_{\Delta} &=& \int d\vec x
\left(
\begin{array}{cc}
\bar\Delta_e & \bar\Delta_h
\end{array}
\right)
\left(
\begin{array}{cc}
\Pi_s^{-1} - u_s& -u_3\\
-u_3 & \Pi_s^{-1}-u_s\\
\end{array}
\right)\left(
\begin{array}{c}
\Delta_e\\
\Delta_h\\
\end{array}
\right) + O(|\Delta_e|^4, |\Delta_h|^4).
\end{eqnarray}
The eigen modes of superconducting fluctuators are given by:
\begin{eqnarray}
L_{\pm,++}^{-1} = \Pi_s^{-1}-(u_s\pm u_3) =\left[\chi^{SC}_{\pm,++}\right]^{-1}.
\end{eqnarray}
Two important limiting cases: i) $u_3 \Pi_s \ll 1$ and $u_s \Pi_s \sim 1$ - superconducting fluctuating modes are almost degenerate and basis $(\bar \Delta_e \bar\Delta_h)^T$ represents correct basis for the multi-mode coupling theory; ii) $u_3 \Pi_s \sim  1$  and $u_s \Pi_s \ll 1$ - superconducting fluctuation modes are spit onto $s_{\pm}$ and $s_{++}$ modes, $T_c$ for $s_{\pm}$ is higher and the new basis
$(\bar \Delta_+ \bar\Delta_-)^T$ describes rotated by $\theta=\pi/4$ old basis $\Delta_\pm=(\Delta_e\pm\Delta_h)/\sqrt{2}$.

The magnetic part of action describing SDW fluctuations is obtained by similar way:\\
i) we introduce the 2-spinor
\begin{eqnarray}
\bar \psi^T=
\left(
\begin{array}{cccc}
\bar f_\uparrow & \bar f_\downarrow &\bar c_\uparrow & \bar c_\downarrow
\end{array}
\right)^T,
\end{eqnarray}
ii) we insert vector bosonic fields describing magnetic fluctuations by means of $\delta$-function
\begin{eqnarray}
\int {\cal D}[\vec m] \delta\left(\vec m - \bar f \vec \sigma c - \bar c \vec \sigma f\right),
\end{eqnarray}
iii) implement the integral representation for the functional $\delta$-function
\begin{eqnarray}
\delta\left(\vec m - \bar f \vec \sigma c - \bar c \vec \sigma f\right)=
\int {\cal D}[\vec n]\exp\left(i\vec n\left[\vec m - \bar f \vec \sigma c - \bar c \vec \sigma f\right]\right).
\end{eqnarray}
The next steps are similar to derivation of superconducting fluctuating part: integrating over Grassmann variables and integrating over $\vec n$ by means of the saddle point approximation . As a result, the magnetic fluctuators are described by the Lagrangian:
\begin{eqnarray}
{\cal L}_{\vec m}=\int d \vec x \left\{\vec m   D_{m}^{-1} \vec m \right\}+ O(\vec m^4),
\end{eqnarray}
with fluctuator
\begin{eqnarray}
D_{m}^{-1} =  \Pi_1^{-1}-(u_1+ u_3)=\left[\chi^{SDW}\right]^{-1}.
\end{eqnarray}
Notice, that we excluded the CDW instability characterized by the fluctuator of CDW mode
$\Sigma \sim ( \bar f_\alpha \delta_{\alpha\beta} c_\beta + c.c. )$:
$D_{\Sigma}^{-1} =  \Pi_1^{-1}-(u_1- u_3)=\left[\chi^{CDW}\right]^{-1}$ due to lower $T_{CDW}< T_{s}$.

Finally, we elaborate on the coupling between superconducting and magnetic modes. For this sake we need a
"super-Nambu" 4-spinor
\begin{eqnarray}
\bar \Upsilon^T=
\left(
\begin{array}{cccccccc}
\bar f_\uparrow & \bar f_\downarrow &\bar c_\uparrow & \bar c_\downarrow &
f_\uparrow & f_\downarrow & c_\uparrow & c_\downarrow
\end{array}
\right)^T.
\end{eqnarray}
Performing a loop expansion (see Fig. \ref{gl_2}) we get
\begin{eqnarray}
{\cal L}_{\vec m \Delta}=\int d \vec x \left\{C_1 \left(|\Delta_e|^2+|\Delta_h|^2\right)
+ C_2 \left(\bar\Delta_e\Delta_h+\bar\Delta_h\Delta_e\right)\right\}\vec m^2,
\end{eqnarray}
where $C_1$ and $C_2$ are defined in the next section. 
\subsection*{Ginzburg-Landau approach}
\label{sec:apc}
In order to establish a correspondence between the microscopic fluctuation theory of competing modes
and the macroscopic thermodynamic description, we can derive an effective Ginzburg-Landau (GL) functional. The GL functional for $\delta {\cal F} = {\cal F}_{\rm ord}-{\cal F}_{\rm norm}$ should be written in terms of order parameters corresponding to competing modes (see, e.g. \cite{Chub2010, Schmal2014, Chub2015a, Chub2015b, Levch2016}).

{\color{black}
We present the GL functional for two important cases discussed in the paper.\\
i) the magnetic fluctuating mode $\vec m(\vec x)$ is coupled to $\Delta_-(\vec x)$ superconducting mode:
\begin{eqnarray}
\delta {\cal F}\left[\Delta_-, \vec m \right] =
\int d \vec x\left\{ c_\Delta \nu|\nabla \Delta_-|^2 + c_m \nu (\nabla \vec m)^2 +\alpha_\Delta|\Delta_-|^2+
\alpha_m \vec m^2
+ \frac{1}{2} A|\Delta_-|^4 + B \vec m^4 +(C_1-C_2)|\Delta_-|^2 \vec m^2+...\right\},\nonumber\\
\end{eqnarray}
ii) the magnetic fluctuating mode $\vec m(\vec x)$ is coupled to two superconducting fluctuationg modes $\Delta_e(\vec x)$  and $\Delta_h(\vec x)$:
\begin{eqnarray}
\delta {\cal F} \left[\Delta_e, \Delta_h, \vec m \right] &=&
\int d \vec x\left\{ c_\Delta \nu \left(|\nabla \Delta_e|^2 + |\nabla \Delta_h|^2\right) + c_m \nu (\nabla \vec m)^2 +\alpha_\Delta \left(|\Delta_e|^2+|\Delta_h|^2\right) +\alpha_m \vec m^2 +\right.\nonumber\\
&& \left. A \left(|\Delta_e|^4+|\Delta_h|^4\right) +B \vec m^4 +C_1\left(|\Delta_e|^2+|\Delta_h|^2\right)\vec m^2+
C_2\left(\bar\Delta_e\Delta_h +\bar\Delta_h\Delta_e\right)\vec m^2+...\right\}.
\end{eqnarray}
The coefficients
$\alpha_\Delta=\ln(T/T_c)$
and
$\alpha_m=\ln(T/T_s)$
change sign at the corresponding transition temperatures (in the case of the transition between the ordered states SDW or SC and the coexistent states, the closed loops are constructed from corresponding anomalous Green functions).  The coefficients $c_\Delta=\xi_s^2$ and $c_m=\xi_m^2$ in front of the gradient terms are obtained in a standard way from the small momentum expansion of the polarization loops $\Pi_s$ and $\Pi_1$, $\xi_s$ and $\xi_m$ are superconducting and magnetic coherence lengths respectively (in the "dirty" limit the coherence length $\xi_{s/m}$ should be replaced by
$\sqrt{l\cdot\xi_{s/m}}$ where $l$ is the mean-free path. This replacement changes the Ginzburg number accordingly), see details in  Ref. \cite{AG59, Sadovskii2006, Varla}.}

The fourth order terms are given by the loops containing four Green's functions (see Fig.\ref{gl_2} and Ref. \cite{Chub2010}).
\begin{eqnarray}
A&=& \Pi_s^{(4)}= \frac{T}{2}\sum _n\int \frac{d^d k}{(2\pi)^d} \left(G^0_e(-k)G^0_e(k)\right)^2
=\frac{T}{2}\sum _n\int \frac{d^d k}{(2\pi)^d} \left(G^0_h(-k)G^0_h(k)\right)^2= \frac{\nu T}{2} \sum_n\int_{-\infty}^\infty \frac{d E}{(E^2+\epsilon_n^2)^2},\\
B&=& \Pi_1^{(4)}= \frac{T}{2}\sum _n\int \frac{d ^3k}{(2\pi)^3} \left(G^0_h(k)G^0_e(k)\right)^2
=\frac{T}{2}\sum _n\int \frac{d^d k}{(2\pi)^d} \left(G^0_e(k)G^0_h(k)\right)^2= \frac{\nu T}{2} \sum_n\int_{-\infty}^\infty \frac{d E}{(E^2+\epsilon_n^2)^2},\\
C_1&=& \Pi_{1s}^{(4)}=  T\sum _n\int \frac{d ^3k}{(2\pi)^3}
G^0_h(k)G^0_h(-k)G^0_e(-k)G^0_h(-k)=
 T\sum _n\int \frac{d^d k}{(2\pi)^d} G^0_e(k)G^0_e(-k)G^0_h(-k)G^0_e(-k),\\
C_2&=& \tilde\Pi_{1s}^{(4)}=\frac{T}{2}\sum _n\int \frac{d^d k}{(2\pi)^d}G^0_e(k)G^0_e(-k)G^0_h(k)G^0_h(-k)=\frac{1}{2}\nu T \sum_n\int_{-\infty}^\infty \frac{d E}{(E^2+\epsilon_n^2)^2},
\end{eqnarray}
and, as a result \cite{Chub2010}
\begin{eqnarray}
A=B=\frac{1}{2}C_1=C_2=\frac{\pi \nu T}{2}\sum_{n>0}\frac{1}{\epsilon_n^3}=\frac{7\zeta(3)\nu}{16\pi^2 T^2}.
\end{eqnarray}

Note that the coefficient in front of the mixed term $\sim |\Delta|^2 \vec m^2$ is different in the
$s_{++}$ ($C_{++}=C_1+C_2=3A$) and $s_\pm$  ($C_{+-}=C_1-C_2=A$) modes \cite{Chub2010}.


\section*{Acknowledgements}
We thank B.L. Altshuler, A. Chubukov, O. Dolgov, B. Keimer, H. Klaus, K. Koepernik  and V. Grinenko for discussions on several aspects of the present work. MK also appreciates discussions of the multi-mode Ginzburg - Landau theory with D. Karki and S. Mandal. The present work was partially supported by the DFG Priority Programme SPP1458, VW Stifftung and the Gradiertenkolleg of the TU Dresden. KA and MK are grateful to IFW Dresden for hospitality.

\section*{Author contributions statement}
{\color{black} S.-L.D and KK contributed to the formulation of the problem. MNK, DVE and KK performed the analytic calculations for the multi-mode coupling theory, S.-L.D, J.v.d.B and KK analysed the phase diagram. All authors contributed to writing and reviewing the manuscript.}

\section*{Additional Information}

The authors declare no competing financial interests.

\end{document}